\documentclass[aps,prd,preprint,tightenlines,showpacs,preprintnumbers,nofootinbib,byrevtex]{revtex4}
\usepackage{amssymb,latexsym}
\usepackage{amsmath,amsbsy}
\usepackage{epsfig,bm}
\begin{document}

\def\a{{\alpha}}
\def\b{{\beta}}
\def\d{{\delta}}
\def\D{{\Delta}}
\def\e{{\varepsilon}}
\def\g{{\gamma}}
\def\G{{\Gamma}}
\def\k{{\kappa}}
\def\l{{\lambda}}
\def\L{{\Lambda}}
\def\m{{\mu}}
\def\n{{\nu}}
\def\o{{\omega}}
\def\O{{\Omega}}
\def\S{{\Sigma}}
\def\s{{\sigma}}
\def\th{{\theta}}
\def\x{{\xi}}
\def\Pperp{{\mathbf{P}^{\perp}}}
\def\Pplus{{P^{+}}}
\def\kperp{{\mathbf{k}^{\perp}}}
\def\kplus{{k^{+}}}
\def\dperp{{\mathbf{\Delta}^\perp}}
\def\kpperp{{\mathbf{k}^{\prime \perp}}}
\def\xpp{{x^{\prime \prime}}}
\def\xp{{x^\prime}}
\def\xt{{\tilde{x}}}
\def\eps{{\varepsilon}}
\def\kppperp{{\mathbf{k}^{\prime \prime \perp}}}
\def\Pp{{P^\prime}}
\def\lp{{\lambda^\prime}}

\def\Dslash{\D\hskip-0.65em /}

\def\Pslash{\ol P\hskip-0.65em /}
\def\lslash{l\hskip-0.35em /}
\def\Pslashe{P\hskip-0.65em /}

\def\ol#1{{\overline{#1}}}

\preprint{NT-UW 04-020}
\title{Double distributions for the proton}
\author{B.~C.~Tiburzi}
\email[]{bctiburz@u.washington.edu}
\author{W.~Detmold}
\email[]{wdetmold@phys.washington.edu}
\author{G.~A.~Miller}
\email[]{miller@phys.washington.edu}
\affiliation{Department of Physics\\  
	University of Washington\\     
	Box 351560\\
	Seattle, WA 98195-1560}
\date{\today}

\begin{abstract}
We derive double distributions for the proton in a simple model that contains scalar as well as axial-vector diquark 
correlations. The model parameters are tuned so that the experimentally measured electromagnetic form factors of the proton 
are reproduced for small momentum transfer. Resulting generalized parton distributions satisfy known constraints, 
including the positivity bounds.
\end{abstract}

\pacs{13.40.Gp, 13.60.Fz, 14.20.Dh}

\maketitle

\section{Introduction}

Considerable attention has been focused on generalized parton distributions 
(GPDs)~\cite{Muller:1994fv,Radyushkin:1996ru,Radyushkin:1996nd,Ji:1997ek,Ji:1997nm}
over the past few years. These functions arise from hadronic matrix elements that are non-diagonal in momentum space and 
contain quark or gluon operators separated by a lightlike distance. Traditional parton distributions, which arise in inclusive 
deep-inelastic scattering, involve diagonal hadronic matrix elements of lightlike separated operators. Elastic form factors, on the other hand, 
are non-diagonal matrix elements of local current operators and are probed in exclusive electromagnetic reactions, for example. 
Thus GPDs encompass physics of both inclusive and exclusive processes. At leading twist, these structure functions 
describe the soft physics that enters in a variety of hard-exclusive reactions,  
see the reviews~\cite{Ji:1998pc,Radyushkin:2000uy,Goeke:2001tz,Belitsky:2001ns,Diehl:2003ny}.

Because lightlike correlations enter into the definitions of GPDs, these distributions  have 
a naturally simple decomposition for spacelike processes in terms of light-cone Fock components of the initial and final hadronic states.
Employing the light-cone gauge, a representation of GPDs in terms of overlaps of light-cone wave functions has been 
derived~\cite{Diehl:2000xz,Brodsky:2000xy}. This wave function representation is ideal for physical intuition because 
of the intrinsic link to many-body quantum mechanics. A physical picture emerges when one relates GPDs to
momentum-dependent distributions of partons in the transverse plane~\cite{Burkardt:2000za,Diehl:2002he,Burkardt:2002hr}.
The wave function representation of GPDs has another salient feature: the positivity 
bounds~\cite{Radyushkin:1998es,Pire:1998nw,Diehl:2000xz,Pobylitsa:2001nt,Pobylitsa:2002gw,Pobylitsa:2002iu,Pobylitsa:2002ru} 
are transparent. Additionally there is the possibility to elucidate the continuity of GPDs in light-cone time-ordered perturbation theory~\cite{Tiburzi:2002sx}.
Despite these virtues, however, little has been done to understand how the reduction properties of GPDs 
arise within the light-cone wave function representation. For example, lack of manifest Lorentz invariance
makes the polynomiality of GPD moments seem miraculous. Consequently the construction of useful models for GPDs based on light-cone wave functions
is severely limited.

In an alternate approach, one formulates the GPDs in terms of underlying double distributions (DDs)~\cite{Radyushkin:1997ki,Radyushkin:1998es}. 
The GPDs are then obtained as projections of the Lorentz invariant DDs. Not surprisingly this formalism explains the polynomiality properties 
of GPD moments.  For this reason, the DD formulation naturally has become attractive to model builders:
DDs are used almost exclusively to model GPDs. 
In spite of this popularity, calculations of model DDs have only recently been pursued~\cite{Mukherjee:2002gb,Pobylitsa:2002vw}. 
A drawback to using models based on DDs is that the positivity bounds become obscure.
Possible marriage of DDs and positivity bounds has been addressed in a particular framework~\cite{Pobylitsa:2002vi}.  
Overall there is still little insight in constructing DDs with dynamical content. 
Furthermore, as we pointed out, connection to light-cone wave 
functions and the intuition of the Fock space expansion is difficult at best~\cite{Tiburzi:2002kr}, 
except in simple model scenarios \cite{Tiburzi:2002tq,Tiburzi:2003ja,Tiburzi:2004ye}.

In this work, we extend the simple scenarios so far pursued to build a DD model for the proton.
We treat the proton as a bound state of a residual quark and two quarks strongly coupled in both
the scalar and axial-vector diquark channels. The resulting light-cone wave function of the 
proton has appropriate spin structure: containing correlations where the residual quark carries the
spin of the proton as well as correlations where the quark and diquark are in a relative $p$-wave.
We use this model to derive DDs for the proton. Inclusion of the spin structure into double distributions 
is crucial if one wishes to make contact with the spin sum rule for hadrons. While this (in essence two-body) 
model for the proton is crude, model parameters can be tuned so that measured electromagnetic form factors are well 
described at small momentum transfer. The quark distributions would need to be matched so that resulting GPDs 
are suitable for phenomenology.

This paper has the following organization. First we review our conventions for DDs and their relation to GPDs in Sec.~\ref{sec:def}. 
We obtain three DD functions for the proton since these quantities are directly encountered in our calculations.
Next in Sec.~\ref{sec:DDcalc}, we present the quark-diquark model 
under consideration.  DDs are calculated in this model in both the scalar and axial-vector diquark channels. Relevant 
identities are gathered in Appendix~\ref{sec:gamma}, while the details of the derivation appear in Appendix~\ref{sec:DDDD}.
Intuition about the model and its construction is provided in Appendix~\ref{sec:LC}, where the effective light-cone wave function
is extracted from projecting onto the light-cone.  Section~\ref{sec:fit} presents phenomenological uses for the model. 
The model is tuned to reproduce the Dirac and Pauli form factors of the proton for small momentum transfer. 
Consequently these simple, model GPDs satisfy known constraints, including positivity. 
A conclusion ends the paper (Sec.~\ref{sec:concl}).

\section{Definitions} \label{sec:def}

To begin, we set forth our conventions for DDs and their relation to GPDs. 
Moments of DDs appear naturally in the decomposition of twist-two operators' matrix elements that are non-diagonal
in momentum space; moreover, 
they provide an elegant explanation of the polynomiality property of GPDs.  
The non-diagonal proton matrix elements of twist-two operators can be decomposed in a fully Lorentz 
covariant fashion in terms of various twist-two form factors $A_{nk}(t)$, $B_{nk}(t)$ and $C_{nk}(t)$, namely
\begin{multline} 
\langle P^\prime,\lp | 
\ol \psi (0) \gamma^{\{\mu}i\tensor D {}^{\mu_1} \cdots i\tensor D {}^{\mu_n\}} \psi(0)
| P,\l \rangle  
 \\ 
= \ol u_{\lp}(\Pp) \gamma^{\{ \mu } u_{\l}(P) 
\sum_{k=0}^{n} \frac{n!}{ k! (n-k)!} A_{nk}(t) 
\ol P {}^{\mu_1} \cdots \ol P {}^{\mu_{n-k}} 
\left( - \frac{\D}{2}\right)^{\mu_{n-k+1}} \cdots \left( - \frac{\D}{2}\right)^{\mu_{n}\}}
\\
+ \ol u_{\lp}(\Pp) \frac{i \sigma^{ \{ \mu \nu } \D_\nu}{2 M}  u_{\l}(P) 
\sum_{k=0}^{n} \frac{n!}{ k! (n-k)!} B_{nk}(t) 
\ol P {}^{\mu_1} \cdots \ol P {}^{\mu_{n-k}} 
\left( - \frac{\D}{2}\right)^{\mu_{n-k+1}} \cdots \left( - \frac{\D}{2}\right)^{\mu_{n}\}}
\\
- \ol u_{\lp}(\Pp) \frac{\D^{ \{ \mu} }{4 M}  u_{\l}(P) 
\sum_{k=0}^{n} \frac{n!}{ k! (n-k)!} C_{nk}(t) 
\ol P {}^{\mu_1} \cdots \ol P {}^{\mu_{n-k}} 
\left( - \frac{\D}{2}\right)^{\mu_{n-k+1}} \cdots \left( - \frac{\D}{2}\right)^{\mu_{n}\}} \label{eqn:moments}
,\end{multline}
where the action of ${}^{\{}\cdots{}^{\}}$ on Lorentz indices produces the symmetric, traceless part of the tensor,
$\ol P$ is defined to be the average momentum between the initial and final states $\ol P {}^\mu = \frac{1}{2} ( P' + P)^\mu$,
and $\D$ is the momentum transfer $\D^\mu = ( P' - P)^\mu$, with $t = \D^2$.  
$T$-invariance restricts $k$ in the first two sums to be even and odd in the last sum. 
There are three Dirac structures in the above decomposition since in general the 
twist-two currents are not conserved, hence a structure proportional to $\D^\mu$ is allowed.
The decomposition above into various form factors is in fact ambiguous.  
Such difficulties in constructing DDs have been addressed in the literature 
\cite{Polyakov:1999gs,Belitsky:2000vk,Teryaev:2001qm,Tiburzi:2002tq},
and we find the construction in Eq.~\eqref{eqn:moments} directly in our calculations. 
The ambiguity of DDs for spin-$\frac{1}{2}$ systems has been addressed in \cite{Tiburzi:2004qr}.

The above decomposition can 
be used to define three double distributions as generating functions for the twist-two form factors
\begin{align} \label{eqn:generate}
A_{nk}(t) &= \int_{-1}^{1} d\b \int_{-1 + |\b|}^{1 - |\b|} d\a 
\b^{n - k} \a^k F(\b,\a;t), \\
B_{nk}(t) &= \int_{-1}^{1} d\b \int_{-1 + |\b|}^{1 - |\b|} d\a 
\b^{n - k} \a^k K(\b,\a;t), \\
C_{nk}(t) &= \int_{-1}^{1} d\b \int_{-1 + |\b|}^{1 - |\b|} d\a 
\b^{n - k} \a^k G(\b,\a;t) \label{eqn:generated}
.\end{align} 
As a consequence of the restriction on $k$ in the sums, the functions $F(\b,\a;t)$ and $K(\b,\a;t)$ are even in $\a$ while
$G(\b,\a;t)$ is odd. The $F(\b,\a;t)$ and $K(\b,\a;t)$ DDs are similar in form to the functions originally employed in \cite{Radyushkin:1997ki}. 
The difference is due to the third DD, $G(\b,\a;t)$, which incorporates the $D$-term \cite{Polyakov:1999gs}, among other things.

Using the operator product expansion, we can relate the moments in Eq.~\eqref{eqn:moments} to matrix elements of
a bilocal operator. By construction, the DD functions appear in the decomposition of the light-like separated 
quark bilinear operator
\begin{multline} \label{eqn:bilocal}
\langle P^\prime,\lp | 
\ol \psi_q \left( - z/2 \right) \rlap \slash z
\psi_q \left( z/2 \right) 
| P,\l \rangle 
=
\int_{-1}^{1} d\b \int_{-1 + |\b|}^{1 - |\b|} d\a \;
e^{ - i \b \ol P \cdot z + i \a \D \cdot z / 2}
\\
\times 
\ol u_{\lp}(\Pp) 
\Bigg[ \rlap \slash z  F_q(\b,\a;t)
+
\frac{i \sigma^{\mu \nu} z_\mu \D_{\nu}}{2 M}
K_q(\b,\a;t) 
-
\frac{\D \cdot z}{4 M} G_q(\b,\a;t)
\Bigg] u_{\l}(P) 
,\end{multline}
where we have appended a flavor subscript $q$ in the relevant places and $z^\mu$ is a lightlike vector, $z^2 = 0$.

Now we define the light-cone correlation function by Fourier transforming with respect to the light-cone separation $z^-$
\begin{equation} \label{eqn:lcc}
\mathcal{M}_q^{\lp,\l}(x,\x,t) = \frac{1}{4 \pi} \int dz^- 
e^{i x \ol P {}^+ z^-}
\langle P^\prime, \lp | 
\ol \psi_q \left( - z^-/2 \right) \gamma^+ 
\psi_q \left( z^-/2 \right) 
| P, \l \rangle
.\end{equation}
Above the variable $\x$, or skewness parameter, is defined by $\x = - \D^+ / (2 \ol P {}^+)$. As is customarily done, 
we assume without loss of generality that $\x > 0$. 
The correlation function in Eq.~\eqref{eqn:lcc} can be written in terms of the two independent GPDs $H_q(x,\x,t)$ and $E_q(x,\x,t)$ as
\begin{equation} \label{eqn:lccor}
\mathcal{M}_q^{\lp,\l}(x,\x,t) = 
\frac{1}{2 \ol P {}^+} \ol u_{\lp}(\Pp) \left[ 
\gamma^+ H_q(x,\x,t)
+ 
\frac{i \sigma^{+\nu} \D_\nu }{2 M}
E_q(x,\x,t)     
\right] u_{\l}(P)   
.\end{equation}
Unlike the DDs, these GPDs are quantities that enter directly into the amplitude for deeply virtual Compton scattering (DVCS), for example.
Inserting the DD decomposition Eq.~\eqref{eqn:bilocal} into the correlator in Eq.~\eqref{eqn:lccor}, we 
can express the GPDs as projections of the DDs
\begin{align}
H_q (x,\x,t) = \int_{-1}^{1} d\b \int_{-1 + |\b|}^{1 - |\b|} d\a \;
\delta(x - \b - \x \a) 
\left[ F_q(\b,\a;t) + \x G_q(\b,\a;t)
\right],
\\
E_q (x,\x,t) = \int_{-1}^{1} d\b \int_{-1 + |\b|}^{1 - |\b|} d\a \;
\delta(x - \b - \x \a) 
\left[ K_q(\b,\a;t) + \x G_q(\b,\a;t)
\right],
\end{align}
from which we can view the $\x$-dependence of GPDs as arising from different slices of Lorentz invariant DDs. 
Due to the symmetry of the DDs with respect to $\a$, the GPDs $H_q(x,\x,t)$ and $E_q(x,\x,t)$ 
are both even functions of the skewness parameter $\x$.

The GPD $H_q(x,\x,t)$ has an important reduction property. Taking the diagonal limit of the light-cone correlator
Eq.~\eqref{eqn:lccor}, we recover the forward parton distributions,
\begin{equation} \label{eqn:quark}
f_q(x) = H_q(x,0,0) = \int_{-1 + |x|}^{1 - |x|} d\a \,  F_q(x,\a;t).
\end{equation}
In DVCS, the relevant current operators produce the charge and flavor structure $\sum_q e_q^2$ since there are two photons
and thus the charge squared weighted GPDs enter in relevant physical amplitudes.
To discover the relation of GPDs to electromagnetic form factors it is advantageous to consider the 
single photon structure $\sum_q e_q$ and define
\begin{align}
H(x,\x,t) & = \sum_q e_q H_q (x,\x,t), \\  
E(x,\x,t) & = \sum_q e_q E_q (x,\x,t)
.\end{align}
Since $G_q(\b,\a;t)$ is an odd function of $\a$, we have $\int_{-1}^{1} d\b \int_{-1 + |\b|}^{1 - |\b|} d\a  \, G_q(\b,\a;t)  = 0$ and consequently the sum rules
\begin{align}
\int_{-1}^{1} dx \, H (x,\x,t) =  \sum_q e_q \int_{-1}^{1} d\b \int_{-1 + |\b|}^{1 - |\b|} d\a  \, F_q(\b,\a;t)  &= F_1 (t), \\
\int_{-1}^{1} dx \, E (x,\x,t) =  \sum_q e_q \int_{-1}^{1} d\b \int_{-1 + |\b|}^{1 - |\b|} d\a  \, K_q(\b,\a;t)  &= F_2 (t),
\end{align}
which relate the zeroth moments of the GPDs to the Dirac and Pauli form factors, $F_{1,2}(t)$.

The GPDs satisfy further constraints arising from the norm on Hilbert space: the positivity bounds 
\cite{Radyushkin:1998es,Pire:1998nw,Diehl:2000xz,Pobylitsa:2001nt,Pobylitsa:2002gw,Pobylitsa:2002iu,Pobylitsa:2002ru} . 
These bounds are particularly important for comparing with experiment. Model GPDs which reduce to the experimental 
quark distribution in Eq.~\eqref{eqn:quark} but violate the positivity bounds should not be considered because one 
knows from the outset that rate estimates predicted by such model GPDs are automatically wrong. Violation of 
the positivity bounds is a signal that the model is inconsistent with the underlying field theory. Perhaps surprisingly,
such violation occurs frequently in many standard hadronic models.
Of interest to us are the basic bounds for both the spin-flip and non-flip amplitudes
\begin{equation} \label{eqn:positivity}
\theta ( x - \x) \Big| \mathcal{M}_q^{\l,\pm \l}(x,\x,t) \Big| \leq 
\sqrt{ f_q\left( \frac{x- \x}{1 - \x}\right) f_q\left( \frac{x + \x}{1 + \x}\right)}
,\end{equation}
which we use as a stipulation in constructing our model.

Lastly we need to address the negative range of the DD variable $\b$. 
Experimentally and diagrammatically $\b$ is strictly positive and crossing
symmetry can be used to relate the functions for positive and negative values of $\b$. 
To this end, we define two functions for each DD
\begin{eqnarray} \label{eqn:Fplus}
F_q^{\pm}(\b,\a;t) & = &  F_q(\b,\a;t) \pm  F_{\ol q} (\b,\a;t), \\
K_q^{\pm}(\b,\a;t) & = &  K_q(\b,\a;t) \pm K_{\ol q} (\b,\a;t),  \\
G_q^{\pm}(\b,\a;t) & = &  G_q(\b,\a;t) \pm  G_{\ol q} (\b,\a;t) \label{eqn:Gplus}  
,\end{eqnarray}
where the antiquark contributions are defined by crossing 
\begin{eqnarray}
F_{\ol q}(\b,\a;t)  &=& - F_{q}(-\b,\a;t),  \notag \\
K_{\ol q}(\b,\a;t) &=& - K_{q}(-\b,\a;t),  \notag \\
 G_{\ol q}(\b,\a;t) &=& - G_{q}(-\b,\a;t) \notag
.\end{eqnarray}
Thus the plus DDs [$F_q^+(\b,\a;t)$, $K_q^+(\b,\a;t)$, and $G_q^+(\b,\a;t)$] are odd functions of $\b$ 
and the minus DDs [$F_q^-(\b,\a;t)$, $K_q^-(\b,\a;t)$, and $G_q^-(\b,\a;t)$] are even functions. 
In partonic language, the minus DDs correspond to 
a difference in quark and antiquark DDs (the valence configuration) while the plus DDs 
are a sum of quark and anitquark DDs (the total configuration).

By virtue of the above definitions Eqs.~\eqref{eqn:Fplus}--\eqref{eqn:Gplus}, we can remove the 
explicit negative-$\b$ parts from DDs and GPDs by defining
\begin{eqnarray}
H^\pm_q(x,\x,t) &=& H_q(x,\x,t) \pm H_{\ol q} (x,\x,t),  \\
E^\pm_q(x,\x,t) &=& E_q(x,\x,t) \pm E_{\ol q} (x,\x,t)
,\end{eqnarray}
where the antiquark contributions are defined by crossing, analogous to the DDs above. 
In this form, we can rewrite the reduction relations in a more familiar way
\begin{equation} \label{eqn:qval}
f^\pm_q(x) = H^\pm_q(x,0,0) = \int_{-1 + x}^{1 - x} d \a \, F^\pm_q(x,\a;t)
,\end{equation}
where the total and valence quark distributions are $f^\pm_q(x) = f_q(x ) \pm f_{\ol q} (x)$.
The zeroth moment sum rules are thus
\begin{align}
\int_{0}^{1} dx \, H^- (x,\x,t) =  \sum_q e_q \int_{0}^{1} d\b \int_{-1 + \b}^{1 - \b} d\a  \, F^-_q(\b,\a;t)  &= F_1 (t), \\
\int_{0}^{1} dx \, E^- (x,\x,t) =  \sum_q e_q \int_{0}^{1} d\b \int_{-1 + \b}^{1 - \b} d\a  \, K^-_q(\b,\a;t)  &= F_2 (t)
,\end{align}
and depend only on the valence configurations. 
In our simple valence model for the proton, we only address quark configurations. The double distribution variable 
$\b$ as well as the momentum fraction $x$ are positive below.
The above positivity bounds in Eq.~\eqref{eqn:positivity} also hold for valence and total distributions and amplitudes 
constructed from the valence and total GPDs.

\section{Model double distributions}\label{sec:DDcalc}

To calculate DDs for the proton, as a first step we use only a simple model consisting of two quarks strongly coupled 
in the scalar and axial-vector diquark channels along with a residual quark. We assume that the residual quark
is a free particle of mass $m$. 
This model can be considered as loosely based on relativistic quark models~\cite{Chung:1991st}
or on the Nambu-Jona Lasinio model of the proton in the static approximation, see, e.g., \cite{Buck:1992wz,Mineo:1999eq}.
We keep full Lorentz covariance in order to preserve the polynomiality of the moments Eq.~\eqref{eqn:moments}. 
Without covariance, once cannot deduce the DDs.

Currently two-loop calculations, in which the diquark structure is resolved, prevent us from obtaining analytic results. 
Thus for tractability, we treat the scalar and axial-vector diquark $T$-matrices as free particle propagators, writing 
\begin{align} \label{eqn:scalar}
D(k) & = \frac{i}{k^2 - m_{SD}^2 + i \varepsilon}, \\
D^{\mu \nu}(k) & =  - i \; \frac{g^{\mu \nu} - k^\mu k^\nu / m_{AD}^2}{k^2 - m_{AD}^2 + i \varepsilon},
\label{eqn:vector}
\end{align}
respectively. The proton Bethe-Salpeter vertex for our model is thus
\begin{equation} \label{eqn:vertex}
\Gamma (k,P) = \frac{1}{\sqrt{2}} \, \chi^{(s)}  D(P-k)\otimes   \Gamma^{(s)}(k,P) 
+  \frac{1}{\sqrt{2}} \, \chi^{(a)}_{\mu,i} \,  D^{\mu \nu}(P - k) \otimes  \Gamma^{(a)}_{\nu,i}(k,P), 
\end{equation}
where the diquark vertices are direct products of spin and isospin factors
\begin{align}
\chi^{(s)} &= \frac{1}{\sqrt{2}} 
(i \gamma_5 C ) \otimes \frac{1}{\sqrt{2}}( i \tau_2 ), \\
\chi^{(a)}_{\mu, i} &= \frac{1}{\sqrt{6}}
( i \gamma_\mu C ) \otimes \frac{1}{\sqrt{6}} (i \tau_i \tau_2)  
,\end{align} 
and we do not append propagators for the first and second quarks. 
For simplicity we choose the quark-diquark vertex functions to be point-like, namely
\begin{align}
\Gamma^{(s)} &= \openone \otimes \openone , \\
\Gamma^{(a)}_{\nu, i} &= \gamma_5 \gamma_\nu  \otimes \tau_i
.\end{align}
This choice corresponds to modeling only a subset of the eight possible structures for the proton wave function.
However in more realistic quark-diquark models, these are the dominant contribution, 
see~\cite{Oettel:1998bk,Oettel:2000jj,Oettel:2000ig} for a complete discussion. The vertex function also contains an overall 
color anti-symmetrization which we suppress throughout. 
The conjugate vertex is given by $\ol \Gamma(k,P) = C \, \Gamma(-k,-P)^{\text{T}} C^\dag$.
One could modify the point-like bound-state vertex with a form factor as is commonly done
in more robust diquark model, but in such models the issue of the positivity bounds is hard to address.

In this model, the axial diquark channel does not contribute to the proton's electromagnetic form factors. Thus we can only determine the 
parameters $m$ and $m_{SD}$ by fits to the Dirac and Pauli form factors. The parameter $m_{AD}$ could be tuned by fitting the quark 
distributions at some scale, however, we shall pursue a simpler course and set $m_{SD} = m_{AD} = m_D$. 
Alternately $m_{AD}$ could be tuned from neutron form factor data.

\begin{figure}
\begin{center}
\epsfig{file=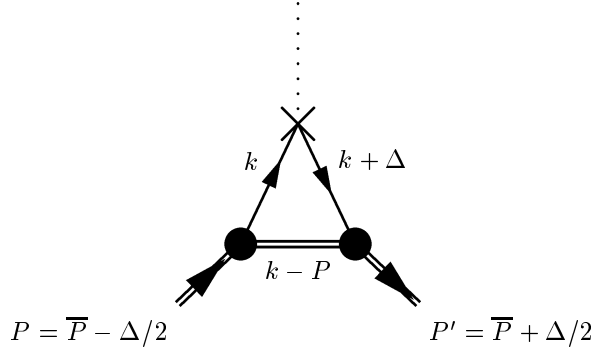}
\caption{Impulse approximation to the twist-two matrix elements. Here the twist-two operators with momentum insertion are denoted
by a cross. The diquark spectator is depicted by a double line and the initial- (final-) state proton has momentum $P$ ($P'$).}
\label{ftwist}
\end{center}
\end{figure}

To derive the $F_q(\b,\a;t)$, $K_q(\b,\a;t)$ and $G_q(\b,\a;t)$ DDs, we consider the action of the twist-two operator 
$\mathcal{O}^{\mu \mu_1 \ldots \mu_n} = \ol \psi_q  \gamma^{\{\mu}i\tensor D {}^{\mu_1} \cdots i\tensor D {}^{\mu_n\}} \psi_q$
between non-diagonal proton states. 
To make any progress in calculating DDs, we must use the parton model simplification for the gauge covariant derivative: 
$D^\mu \to \partial^\mu$ and write 
$\tensor D {}^{\mu} = \frac{1}{2}( \overset{\rightarrow}{\partial}{}^\mu - \overset{\leftarrow}{\partial}{}^\mu )$.
Working in the impulse approximation (see Fig. \ref{ftwist}), 
we have the contributions from both the scalar and axial-vector diquark channels. 
In the scalar diquark channel, we have
\begin{equation} \label{eqn:impulse}
\int  d^4 k 
\frac{\ol u_{\l'} (\Pp) (\rlap\slash k +  \Dslash + m) \Gamma^{\mu \mu_1 \ldots \mu_n} (\rlap\slash k + m ) u_\l(P)}
{[k^2 - m^2 + i \varepsilon] \, [(k+\D)^2 - m^2 + i \varepsilon] \, [(k - \ol P + \D/2)^2 - m_D^2 + i \varepsilon]}
.\end{equation}
The symmetric, traceless tensor $\Gamma^{\mu \mu_1 \ldots \mu_n}$ 
in Eq.~\eqref{eqn:impulse} arises from the momentum space transcription of the parton model operator
$\mathcal{O}^{\mu \mu_1 \ldots \mu_n}$ and is given by 
\begin{equation} \label{eqn:GAMMA}
\Gamma^{\mu \mu_1 \ldots \mu_n} = \gamma^{\{ \mu} (k + \D/2)^{\mu_1} \cdots (k + \D/2)^{\mu_n \}}
.\end{equation}
To regulate\footnote{%
Alternate schemes using Pauli-Villars subtractions often regulate such models
and are also attractive from the perspective of DDs since Lorentz covariance is maintained. These subtractions, however, generally
violate the bounds in Eq.~\eqref{eqn:positivity}. For example, in the NJL model of the pion with two subtractions \cite{Theussl:2002xp}, 
the positivity bounds, which were ignored by the authors, are violated for small values of $-t$. For the case of a quark-diquark model 
regularized via Pauli-Villars subtractions, the violations are more 
severe and persist for all values of $-t$ due to the mismatch of end-point and crossover behavior. This commonly encountered
problem is discussed in \cite{Tiburzi:2002kr}.
For these reasons, we have used the smeared vertex regularization above, 
which we also employed previously for the pion \cite{Tiburzi:2002tq}.
As shown in \cite{Tiburzi:2004ye}, one can employ a very similar analytic regularization
\begin{equation} \label{eqn:smear2}
\gamma^\mu \to \left[ \frac{m^2}{k^2 - m^2 + i \varepsilon} \right]^a \, \gamma^\mu \, 
\left[ \frac{m^2}{(k+ \D)^2 - m^2 + i \varepsilon} \right]^a
,\end{equation}
and still maintain the positivity bounds. As with the smeared vertex, this regularization corresponds to 
modification of the active quark's propagators but leaves the spectator diquark propagator untouched. 
There are a number of drawbacks to this regularization when compared to the smeared vertex. 
For $a>1$, the model form factors fall off too quickly as a function of $-t$. Setting $a=1$, the proton's electromagnetic form 
factors can only be matched at the $5 - 10 \%$ level. Additionally the model double distibutions derived using the analytic
regularization vanish at the boundaries of support, which need not be the case \cite{Tiburzi:2004qr} and is not so with the smeared 
vertex.
}
the above expression, we choose to smear the vertex in the following covariant manner \cite{Bakker:2000pk}
\begin{equation} \label{eqn:smear}
\gamma^\mu \to \frac{\Lambda^2}{k^2 - \Lambda^2 + i \varepsilon} \, \gamma^\mu \, \frac{\Lambda^2}{(k+ \D)^2 - \Lambda^2 + i \varepsilon}
.\end{equation}
Since the NJL model (to which our model bears a resemblance) is non-renormalizable, the choice of scheme is incorporated into the dynamics and hence
the choice of regularization should maintain desired properties. For phenomenological estimates of GPDs, the positivity
bounds are of crucial importance, and our regularization choice respects these bounds, see Appendix~\ref{sec:LC} for details.
One can view the regularization $\Lambda < \infty$ as mimicking the non-local, non-perturbative
structure of the twist-two, quark-antiquark vertex. 
This choice of smearing maintains current conservation but does not respect the Ward-Takahashi identity.\footnote{%
Since the Ward-Takahashi identities only constrain one generalized form factor in Eq.~\eqref{eqn:moments} 
(i.e.~the electromagnetic coupling)
and positivity constrains all generalized form factors non-trivially, we prefer to violate the former and preserve the latter.
Adding the factor $Z(\Lambda)$ to the axial diquark contribution allows us to gauge the extent of violation of the Ward-Takahashi
identity. Moreover, this factor gives the situation a remedy. On the other hand, if the positivity bounds are not respected in 
the construction of the model, it is far from clear how to remedy the violation. 
} Thus the normalization of 
amplitudes in Eq.~\eqref{eqn:vertex} is only approximately preserved. 
The unregularized model is set up so that if the quantities were finite, then the normalization of the Dirac 
form factor $F_1(0) = 1$ implies the correct quark content of the proton, $N_u = 2$, and $N_d = 1$. 
In the smeared vertex regularization, both of these quantities 
acquire $\Lambda$ dependence.  To remedy this feature, we keep the scalar diquark contribution normalized so that $F_1(0) = 1$ 
and add a $\Lambda$-dependent prefactor $Z(\Lambda)$ to the axial diquark contribution. This factor is then adjusted so that 
$N_d = 1$ and consequently the u-quark distribution is correctly normalized, $N_u = 2$. 
An additional drawback of the regularization scheme is that the induced quark form factors do not become
point-like for large momenta. While this is inconsistent with asymptotic freedom, the model is meant only for use in the low 
momentum transfer region where the diquark substructure is not resolved.

The involved numerator structure in Eq.~\eqref{eqn:impulse} complicates calculation of the symmetric traceless part of the tensor, but,  
this can be calculated directly without recourse to explicitly writing out such tensors of rank $n$. The $k^\mu k^\nu$ term of the vector 
propagator leads to exceedingly complicated contributions to DDs that are not in the spirit of our simple model.
Thus we present results including only the $g^{\mu \nu}$ term in the main text, leaving the 
$k^\mu k^\nu$ piece to Appendix~\ref{sec:DDDD}.  The tricks employed to obtain these complicated contributions 
may be useful beyond this work. The numerical effects of this diagonal approximation (where the $k^\mu k^\nu$ term is neglected)
are expected to be small, especially at low momentum transfer \cite{Oettel:1998bk,Oettel:2000jj}.

In order to compactly write out the DDs, we 
define the auxiliary functions
\begin{equation}
D^{ab}(\b,\a;t)^{-1} = \b m_D^2 + \frac{a^2}{2} ( 1 - \b - \a )  + \frac{b^2}{2} ( 1 - \b + \a)  - \b (1-\b) M^2 - [(1-\b)^2 - \a^2] t / 4 
,\end{equation}
which is the typical energy denominator in both channels and
\begin{equation}
\Lambda_{ab} = \frac{\Lambda^4}{(\Lambda^2 - m^2)^2} ( \delta^m_a \delta^m_b  - \delta^m_a \delta^\Lambda_b - \delta^\Lambda_b \delta^m_b 
+ \delta^\Lambda_a \delta^\Lambda_b)
,\end{equation}
which is the typical regularization prefactor for all DDs in this model. 
Although $m$ and $\L$ are not discrete variables, we have employed Kronecker deltas 
as convenient shorthands for fixing the values of $a$ and $b$. 
Notice the contraction $\Lambda_{ab} D^{ab}(\b,\a;t)$ is an even function of $\a$.

Calculation of the DDs for the scalar diquark component of the proton yields
\begin{equation} \label{eqn:Fs}
\begin{pmatrix}
F^{(s)}_q(\b,\a;t) \\
K^{(s)}_q(\b,\a;t) \\
G^{(s)}_q(\b,\a;t)
\end{pmatrix}
= 
N \delta_{q,u}  \Lambda_{ab} 
\left[
\log D^{ab}
\begin{pmatrix}
1 \\
0 \\ 
0
\end{pmatrix}
+
D^{ab}
\begin{pmatrix}
(m + \b M )^2 +  \left[ (1-\b)^2 - \a^2 \right] \frac{t}{4} \\
2 M (1 - \b) ( m + \b M) \\
4 M  \a  ( m + \b M) 
\end{pmatrix}
\right]
,\end{equation}
where $N$ represents the overall normalization which is fixed by the charge of the proton
and $D^{ab}$ is merely an abbreviation for $D^{ab}(\b,\a;t)$.
Despite the behavior of the induced constituent quark form factors, the asymptotics for $F_1(t)$ and
$F_2(t)$ that result from Eq.~\eqref{eqn:Fs} are (up to logs) $1/t^2$ and $1/t^3$, respectively. 
The difference can be traced directly to the orbital angular momentum content of the scalar diquark
light-cone wave function, see Appendix \ref{sec:LC}.

In the the axial-vector diquark channel (keeping only the $g^{\mu \nu}$ term), we have the following DDs 
\begin{multline} 
\begin{pmatrix}
F^{(a)}_q(\b,\a;t) \\
K^{(a)}_q(\b,\a;t) \\
G^{(a)}_q(\b,\a;t)
\end{pmatrix}
= 
\frac{2}{9} Z(\Lambda) \, N \, (2 \delta_{q,d} + \delta_{q,u}) \,  \Lambda_{ab} 
\\
\label{eqn:Fa} 
\times
\left[
\log D^{ab}
\begin{pmatrix}
1 \\
0 \\ 
0
\end{pmatrix} 
+ 
D^{ab}
\begin{pmatrix}
(2 m + \b M )^2  - 3 m^2 -  \left[ (1-\b)^2 - \a^2 \right] \frac{t}{4} \\
2 M \b [ 2 m - (1 - \b ) M] \\
- 4 M  \a  ( 2 m + \b M) 
\end{pmatrix}
\right]
,\end{multline}
where $Z(\Lambda)$ is the regularization dependent factor akin to wave function renormalization. 
As commented above, the value of $Z(\Lambda)$ is
chosen to preserve $N_d = 1$. The full DDs are then given by the sum of the scalar and axial-vector pieces, 
e.g., $F_q(\b,\a;t) = F_q^{(s)}(\b,\a;t) + F_q^{(a)}(\b,\a;t)$.

\section{Phenomenological applications} \label{sec:fit}

As commented above, our philosophy is to 
tune the parameters $m$ and $m_D$ so that proton electromagnetic 
form factor data at low momentum transfer are reproduced. This is particularly simple, 
since the axial diquark does not contribute to these quantities.
For the electromagnetic form factors of the proton, there is high precision data from Jefferson Lab~\cite{Jones:1999rz,Gayou:2001qd} 
and a recent global analysis and parametrization of~\cite{Arrington:2003qk}. The Sachs form factors are known 
experimentally to about $2\%$ accuracy in the small momentum transfer regime and are given by
\begin{align}
G_E(t) & = F_1(t) + \frac{t}{4 M^2} F_2(t), \\
G_M(t) & = F_1(t) + F_2(t)
.\end{align}

\begin{figure}
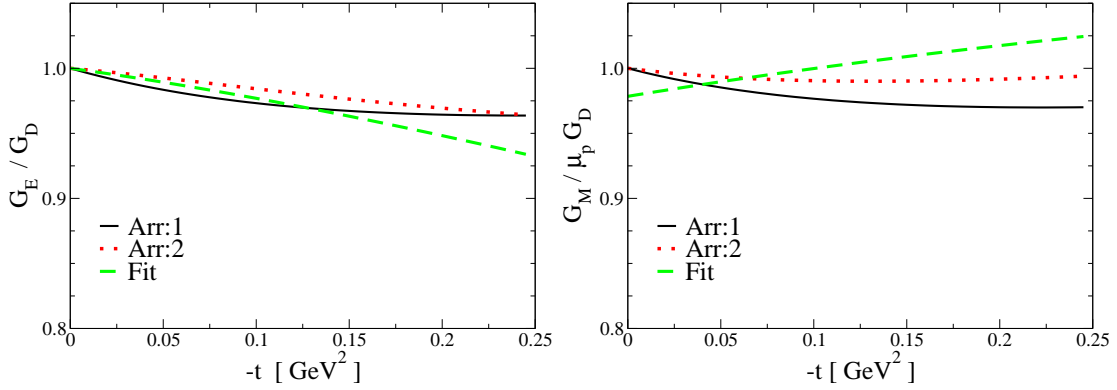

\begin{center}
\epsfig{file=GE.eps,height=2in}
\epsfig{file=GM.eps,height=2in}
\caption{Comparison of fits for $G_E(t)$ and $G_M(t)$ with empirical parameterizations. 
The ratios $G_E(t)/G_D(t)$ and $G_M(t)/ \mu_p G_D(t)$ are plotted
against $-t$ in $\texttt{GeV}^2$. The curves ``Arr:1'' and ``Arr:2'' correspond to the 
parameterizations of $G_E(t)$ and $G_M(t)$ given in Tables I and II of \cite{Arrington:2003qk}, respectively. 
}
\label{fGE}
\end{center}
\smallskip
\end{figure}

The form factor data are reasonably fit (to $\sim 2 - 5 \%$) by the model for $m = 0.445 \texttt{ GeV}$, 
$\L = 0.465 \texttt{ GeV}$, and $m_D = 0.720 \texttt{ GeV}$. For these parameter values the 
normalization $N = 1.52$.  
In Fig.~\ref{fGE}, we compare this phenomenological form factor fit to 
the two parameterizations of~\cite{Arrington:2003qk}. 
As is standard, we plot ratios of electric and magnetic form factors to the empirical dipole form factor, namely
\begin{equation}
G_D(t) = \left(1 - \frac{t}{ M_D^2}\right)^{-2}
,\end{equation}
where the dipole mass squared is $M_D^2 = 0.71 \texttt{ GeV}^2$.

\begin{figure}
\begin{center}
\epsfig{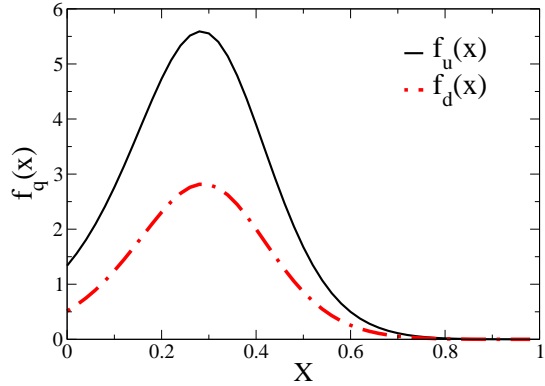}
\caption{Quark distributions for the proton model. The $u$- and $d$-quark distributions are plotted as a function 
of $x$. 
}
\label{fquark}
\end{center}
\end{figure}

\begin{figure}
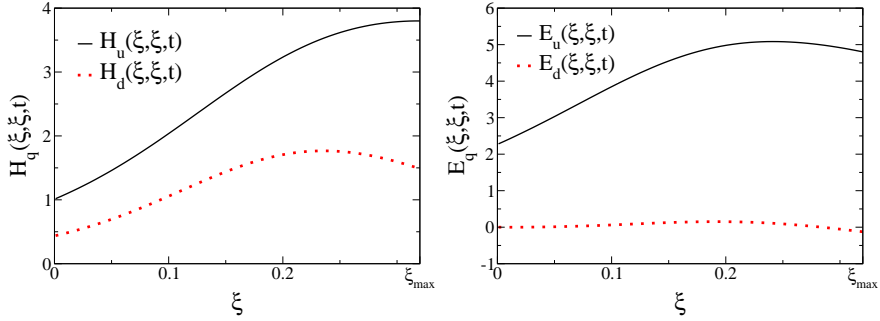

\begin{center}
\epsfig{file=crossH.eps,height=1.65in}
\epsfig{file=crossE.eps,height=1.65in} 
\caption{Proton model GPDs at the crossover. The $u$- and $d$-quark GPDs, $H_{u,d}(x,\x,t)$ and $E_{u,d}(x,\x,t)$, 
are plotted as a function of $\x$ at the crossover $x = \x$ 
for $t = -0.1 \texttt{ GeV}^2$. For this value of $t$, $\x_{\text{max}} = 0.33$.}
\label{fcross}
\end{center}
\end{figure}

We can also determine the $u$- and $d$-quark distributions in our model. Since we do not have antiquarks,
the plus and minus distributions are identical $f^\pm_q(x) = f_q(x)$. In Fig.~\ref{fquark}, we
plot the $u$- and $d$-quark distributions as a function of $x$.
The distributions are properly normalized 
so that $N_u = 2$ and $N_d = 1$. As commented above this normalization requires a relative $\Lambda$-dependent factor
for the axial-diquark contributions, $Z(\Lambda = 0.465 \texttt{ GeV}) = 1.09$. 
Again this is required because the regularization scheme we have chosen 
does not preserve the Ward-Takahashi identity. Without the extra factor, the violation is $\sim 5$--$10 \%$.  
Notice that the distributions do not vanish at the end-point $x = 0$. This is typical of NJL type model calculations. The point-like 
kernel is independent of momentum and hence, when one writes down the effective wave equation for the proton wave function, 
one easily deduces that the wave function should be non-zero at both end-points. This is true of the unregularized wave equation. 
The fact that the model distributions vanish at $x=1$ is due to our choice of regularization. 
Physically it is thus reasonable to think of the regularization as mimicking contributions 
from higher Fock states. The small-$x$ behavior of the model quark distributions, however, compares poorly with experiment.   
\begin{figure}
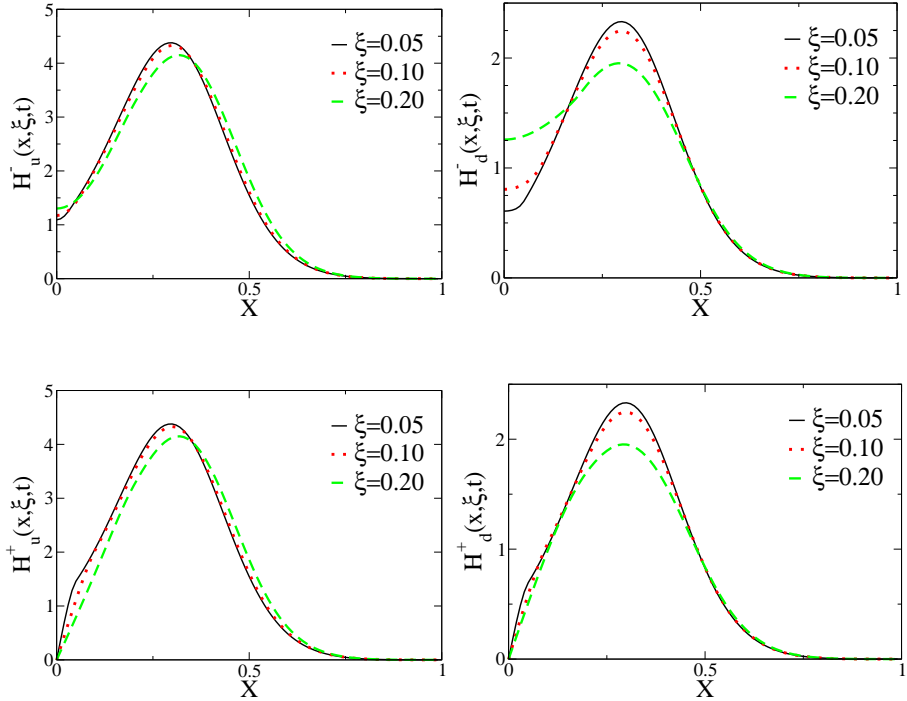

\begin{center}
\epsfig{file=hvalu.eps,height=1.65in}
\epsfig{file=hvald.eps,height=1.65in} \\
\bigskip 
\smallskip
\smallskip
\smallskip
\smallskip
\epsfig{file=hsingu.eps,height=1.65in}
\epsfig{file=hsingd.eps,height=1.65in}
\caption{GPDs for the proton model. The $u$- and $d$-quark GPDs $H^\pm_{u,d}(x,\x,t)$ are plotted as a function 
of $x$ for a few values of $\x$ at $t = -0.1 \texttt{ GeV}^2$. Notice the scale of the $d$-distributions is half that of the $u$.}
\label{fHbare}
\end{center}
\end{figure}
\begin{figure}
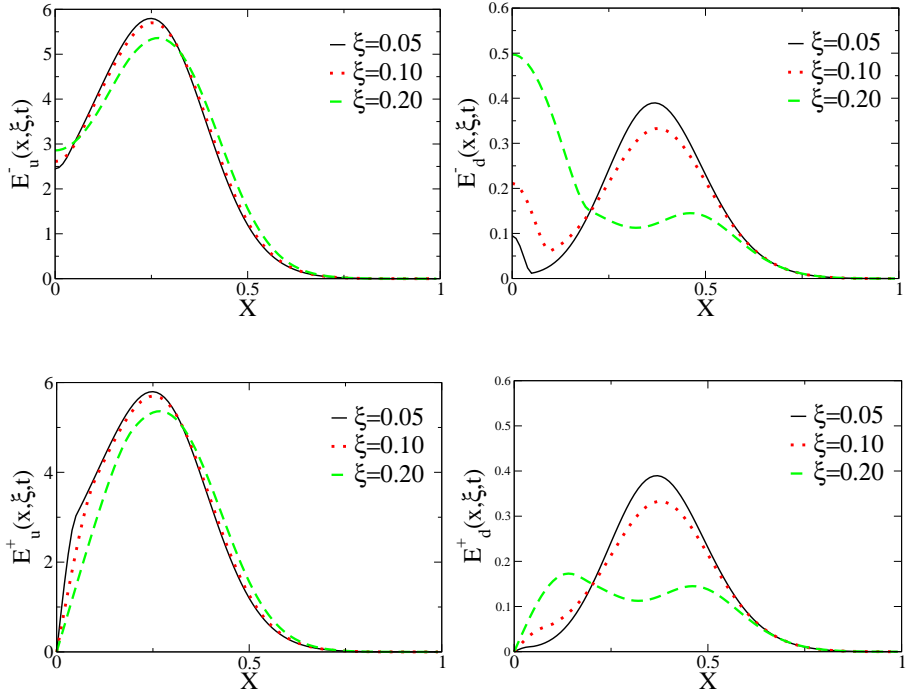

\begin{center}
\epsfig{file=evalu.eps,height=1.65in}
\epsfig{file=evald.eps,height=1.65in} \\
\bigskip 
\smallskip
\smallskip
\smallskip
\epsfig{file=esingu.eps,height=1.65in}
\epsfig{file=esingd.eps,height=1.65in}
\caption{GPDs for the proton model. The $u$- and $d$-quark GPDs $E^\pm_{u,d}(x,\x,t)$ are plotted as a function 
of $x$ for a few values of $\x$ at $t = -0.1 \texttt{ GeV}^2$. Notice the scale of the $d$-distributions is one tenth that of the $u$.}
\label{fEbare}
\end{center}
\end{figure}
Additionally in Figs.~\ref{fHbare} and \ref{fEbare}, we have plotted the $H^\pm_{u,d}(x,\x,t)$ and $E^\pm_{u,d}(x,\x,t)$ GPDs. 
The figures show the GPDs at fixed $-t = 0.1 \texttt{ GeV}^2$ for a few values of $\x$. The value of the $u$- and $d$-quark 
GPDs, $H_{u,d}(x,\x,t)$ and $E_{u,d}(x,\x,t)$, at the crossover $x = \x$ are plotted as a function of $\x$ in Fig.~\ref{fcross} for the
same value of $-t = 0.1 \texttt{ GeV}^2$ .
The distributions plotted in Figs.~\ref{fquark}--\ref{fcross} are presumably at some low hadronic scale intrinsic to the model. 
One way to determine the scale of the model is to use the evolution equations to evolve empirical parameterizations
down to a scale where the first few moments of our model distributions agree. This procedure is not unique; many models can reproduce
the empirical quark distributions at higher scales. Also the use of perturbative evolution is questionable at best at low scales. 
While the evolution kernels for GPDs are known at next-to-leading order \cite{Belitsky:1998gc,Belitsky:1998uk,Belitsky:1999gu,Belitsky:1999fu,Belitsky:1999hf}, 
we caution this approach for our simple model. Perturbative evolution cannot generate the non-perturbative small-$x$ physics which our model lacks
and the small-$x$ physics is crucial for relating to DVCS data. The leading-twist DVCS amplitude is convolution of GPDs and a hard scattering kernel 
that emphasizes a region where the final-state wave function is evaluated near the end-point. In fact, the imaginary part of the amplitude
is directly proportional to an overlap of light-cone wave functions, where the final state has $x=0$.
In order to be useful for experimental comparison, a more pragmatic 
solution would be to augment the model with a realistic parametrization of the valence quark distributions.
This could be done by implanting the realistic distribution via factorization of the $\b$-dependence of the DDs \cite{Mukherjee:2002gb}. 
This choice, while unrealistic for the DDs, may indeed 
be less problematic for GPDs since there is ample allowance for interplay of $x$, $\x$, and $t$ dependence.
From the perspective of model building, one would be merely satisfying the experimental constraints 
by matching form factors and valence quark distributions; whether
or not the resulting GPD models are suitable to make contact with data remains to be seen.

\section{Conclusion} \label{sec:concl}

Above we have used a simple quark-diquark model to calculate DDs for the proton. The model consists of two quarks
strongly coupled in the scalar and axial diquark channels along with a residual quark. The simplicity 
of such a model allows for analytic computation of the DD functions which contain appropriate spin 
structure. The simplicity also allows for us to make contact with the light-cone wave functions and the 
equivalent overlap representation of GPDs. This toy model study thus enables a comparison between 
the physical intuition of the light-cone Fock space representation of GPDs and the manifestly Lorentz invariant
decomposition of DD functions. 

We were careful to choose a regularization scheme that allows for analytic determination of the DDs and 
satisfies both Lorentz invariance and
the positivity bounds required of GPDs. Our model, although toy-like in nature, is used to match 
the electromagnetic form factors of the proton and can be augmented with realistic valence quark distributions
or evolved up from its intrinsic scale. 

The scope of our continuing investigation is two fold. One direction is to calculate as many DDs in simple scenarios
as possible. This will give modelers a better sense of the form and behavior of DDs and may assist with empirical 
parameterizations of GPDs. The model used here can be easily extended to calculate double distributions for both the axial and tensor GPDs.
In another nearly orthogonal direction, one can improve upon the proton model used
here in order to see how various features of a realistic proton wave function manifest themselves in processes like DVCS. 
Proper treatment of the three-body nature of the proton requires two-loop calculations. Suitable
analytical and computational tools would have to be developed in order to extract double distributions in such a case. 
The model at hand, however, has qualitative similarities to the three-quark structure of the proton  
and should be considered a good starting point.

\begin{acknowledgments}
This work was funded by the U.~S.~Department of Energy grant DE-FG$03-97$ER$41014$.  
WD is grateful for discussions with R.~Alkofer and M.~Oettel. 
\end{acknowledgments}

\appendix

\section{Basic identities}\label{sec:gamma}

For reference we list identities used in computing DDs above. 
To calculate numerator structures involving the
symmetric and traceless tensors, we used generalized Gordon identities
\cite{Brown:1992db} of the form
\begin{equation} \label{eqn:GGordon}
\ol u_{\l'}(\Pp) \Gamma \, u_\l(P) = \frac{1}{4M} \ol u_{\l'} (\Pp) \left( 2 \{ \Pslash, \Gamma \} + [ \Dslash, \Gamma] \right) u_{\l}(P)
,\end{equation}
where $\Gamma$ is any Dirac matrix, $\ol P {}^\mu = \frac{1}{2} ( \Pp + P)^\mu$ and $\D^\mu = ( P^{\prime} - P)^\mu$.
The usual Gordon identity is a special case of Eq.~\eqref{eqn:GGordon}, namely for $\Gamma = \gamma^\mu$
\begin{equation}
\ol u_{\l'}(\Pp) \gamma^\mu u_{\l}(P)= \frac{1}{2 M} \ol u_{\l'}(\Pp) \left( 2 \ol P {}^\mu + i \sigma^{\mu \nu} \D_\nu  \right)  u_{\l}(P)
.\end{equation}
We also require the following two cases of the general identity. With $\gamma_5 = i \gamma^0 \gamma^1 \gamma^2 \gamma^3$
and $\varepsilon^{0123} = +1$, we have
\begin{align}
\ol u_{\l'}(\Pp) \gamma^\mu \gamma_5 \, u_{\l}(P) 
& = \frac{1}{2 M}  \ol u_{\l'}(\Pp) 
\left(  
\D^\mu \gamma_5 
 - \varepsilon^{\mu \nu \a \b} \sigma_{\a \b} \ol P_\nu  
\right)  u_{\l}(P),  \\
\ol u_{\l'}(\Pp)\sigma^{\mu \nu} u_{\l}(P) 
& = \frac{i}{2 M}  \ol u_{\l'}(\Pp) 
\left( 
\gamma^{\nu} \D^\mu - \gamma^\mu \D^\nu 
+ 2 i  \varepsilon^{\mu \nu \a \b} \ol P_\a \gamma_\b  \gamma_5 
\right)  u_{\l}(P)
.\end{align}


\section{Calculations on the light-cone}\label{sec:LC}

Here we include our conventions for projecting quantities onto the light cone. While the development and derivations
above rely exclusively on manifest Lorentz invariance, the light-cone Fock representation provides transparent 
physical intuition about our model. The light-cone wave functions for the model are admittedly simple and thus provide
a useful guide to understanding the DDs constructed above.

For any Lorentz vector $a^\mu$, we define the light cone coordinates
\begin{equation}
a^\pm = \frac{1}{\sqrt{2}} \left( a^0 \pm a^3 \right)
.\end{equation}
The light-cone spinor $u_{\l}(k,m)$ satisfies the Dirac equation $(\rlap\slash k - m) u_{\l}(k,m) = 0$
and is explicitly given by (e.g., see \cite{Heinzl:1998kz})
\begin{equation} \label{eqn:lcspinor}
u_{\l}(k,m) = \frac{1}{2^{1/4}\sqrt{k^+}} 
\left( \sqrt{2} k^+ + \beta m + \bm{\a}^\perp \cdot \mathbf{k}^\perp  \right) X_{\l}
,\end{equation}
where $\b = \gamma^0$, $\bm{\a} = \gamma^0 \bm{\gamma}$, and the unit spinors $X_\l$ are given by
\begin{align}
X_\uparrow^\dagger & = \frac{1}{\sqrt{2}} \left( 1,0,1,0 \right), \notag \\
X_\downarrow^\dagger & = \frac{1}{\sqrt{2}} \left( 0,1,0,-1 \right) \notag
.\end{align}
Using Eq.~\eqref{eqn:lcspinor}, we derive the useful product of spinors of different mass, momentum, and spin
\begin{equation} \label{eqn:spinorprod}
\ol u_{\lp}(k,m) u_\l(P,M) = \frac{1}{\sqrt{k^+ P^+}} 
\left[ 
\delta_{\lp,\l} (k^+ M + P^+ m) 
- \l \, \delta_{\lp,-\l} ( k^+ P_\l - P^+ k_\l)
\right]
,\end{equation}
where the notation $a_\l = a^1 + i \l a^2$ has been employed with the understanding 
that spins correspond to the signs $\uparrow = +1$ and $\downarrow = -1$.

Using the light-cone spinors, one can find the $H_q(x,\x,t)$ and $E_q(x,\x,t)$ GPDs 
in terms of the light-cone, non-diagonal matrix element $\mathcal{M}_q^{\lp,\l}(x,\x,t)$, namely
\begin{align}
\mathcal{M}_q^{\l,\l}(x,\x,t) & = \frac{1}{\sqrt{1 - \x^2}} \left[ (1 - \x^2) H_q(x,\x,t) - \x^2 E_q(x,\x,t) \right], \\
\mathcal{M}_q^{\l,-\l}(x,\x,t) & =  - \frac{ \l \D_\l }{2 M \sqrt{1-\x^2}}  \, E_q(x,\x,t)
.\end{align} 
These expressions enable an alternate means to derive GPDs for the model considered in Sec.~\ref{sec:DDcalc}.
In this approach, one directly inserts the quark bilocal operator between non-diagonal proton states and 
integrates over the relative light-cone energy $k^-$ at the cost of sacrificing manifest Lorentz
invariance. This description in terms of light-cone Fock components, however, is more intuitive than the DD
formulation. Moreover, the light-cone energy integration clarifies the positivity properties of our model GPDs. 
We shall not present complete expressions for the GPDs on the light-cone, however, the diagonal overlap
region $x>\xi$ is particularly simple to consider and thus we provide the details for the scalar diquark. 
Analysis in the other region, $x<\x$, is similar but the expressions are more cumbersome.

For our model, we can find the lowest Fock component's 
light-cone wave function by projecting the covariant Bethe-Salpeter wave function $\Psi(k,P)$ onto the 
null surface $z^+ = 0$, namely
\begin{eqnarray} 
\psi_{\text{LC}} (x,\mathbf{k}^\perp_{\text{rel}}; s_i,\l) &=& \frac{1}{(2 P^+)^2} \int \frac{d k^-}{2 \pi}
\frac{\ol u_{s_1} (P-k,m)}{\sqrt{1-x}} 
\Big( \gamma^+ i \gamma_5 C \Big) 
\ol u_{s_2}(P-k,m)^{\text{T}} \notag \\ 
&& \times \frac{\ol u_{s_3} (k,m) }{\sqrt{x}} \gamma^+ \Psi(k,P) u_\l(P,M) 
,\label{eqn:lcwfn} \end{eqnarray}
where $x$ is the fraction of the proton's longitudinal momentum carried by the residual quark $(x = k^+ / P^+)$, 
and the relative transverse momentum is $\mathbf{k}^\perp_{\text{rel}} = \mathbf{k}^\perp - x \mathbf{P}^\perp$. 
Above, $\l$ labels the spin of the proton, whereas the $s_i$ label the spins of the three quarks.
We have omitted the color and isospin parts of the wave function 
which are trivial: $\propto \varepsilon_{c_1,c_2,c_3} (\delta_{1,u} \delta_{2,d} - 
\delta_{1,d} \delta_{2,u} ) \delta_{3,u}$ for the scalar diquark.
In the scalar channel, the quark-diquark Bethe-Salpeter wave function is
\begin{equation}
\Psi(k,P) = \frac{i}{\rlap \slash k  - m + i \varepsilon}  \Big ( - i g^{(s)} \Big)  D(P - k)
,\end{equation} 
and above we have chosen a single Dirac component $\Gamma^{(s)} = - i g^{(s)}$ and the coupling constant $g$ has been 
absorbed into the overall normalization. Our choice of vertex functions corresponds to only modeling a subset 
of the possible three quark light-cone wave 
functions of the proton \cite{Ji:2002xn}. Carrying out the projection in Eq.~\eqref{eqn:lcwfn}, yields
\begin{equation} \label{eqn:psi}
\psi_{\text{LC}} (x,\mathbf{k}^\perp_{\text{rel}}; s_i,\l) \propto 
\frac{ s_1 \delta_{s_1,-s_2} [\delta_{s_3,\l} (x M + m ) + \l \, \delta_{s_3, -\l} k_{\text{rel},\l}]}
{x \sqrt{ 1 - x} \left[M^2 - \frac{\mathbf{k}^\perp_{\text{rel}} {}^2 }{x (1-x)} - \frac{m^2}{x} - \frac{m_D^2}{1 -x}\right]}
.\end{equation}
The smeared vertex regularization, Eq.~\eqref{eqn:smear}, 
generates an effective wave function $\psi_{\text{LC}}^{\text{eff}} (x,\mathbf{k}^\perp_{\text{rel}}; s_i,\l)$ 
because the spectator diquark's energy pole is unaffected, namely 
\begin{equation} \label{eqn:psieff}
\psi^{\text{eff}}_{\text{LC}} (x,\mathbf{k}^\perp_{\text{rel}}; s_i,\l) \propto 
\Lambda_a \frac{ s_1 \delta_{s_1,-s_2} [\delta_{s_3,\l} (x M + m ) + \l \, \delta_{s_3, -\l} k_{\text{rel},\l}]}
{x \sqrt{ 1 - x} \left[M^2 - \frac{\mathbf{k}^\perp_{\text{rel}} {}^2 }{x (1-x)} - \frac{a^2}{x} - \frac{m_D^2}{1 -x}\right]}
,\end{equation}
where the function
\begin{equation}
\Lambda_a = \frac{\Lambda^2}{\Lambda^2 - m^2} (\delta_a^m - \delta_a^\Lambda),
\end{equation}
denotes the two terms induced by the regularization scheme. 
In going from Eq.~\eqref{eqn:psi} to Eq.~\eqref{eqn:psieff}, the effect of regularization is to mimic the contribution 
from higher Fock states.
However, there are no true higher Fock components in this model since the interaction kernel is instantaneous in light-cone 
time \cite{Tiburzi:2002sw}.
Focusing on the spin structure, the scalar diquark channel consists of two states: a state 
where the proton spin is aligned with the residual quark's spin and a state where the quark and diquark are in a relative
$p$-wave. Notice the regularization Eq.~\eqref{eqn:smear} does not alter the spin structure of the wave function.

The quark distribution function can be obtained from 
\begin{equation}
f_u(x) = \sum_{s_i} \int d \mathbf{k}^\perp \left| \psi_{\text{LC}}^{\text{eff}} (x,\mathbf{k}^\perp; s_i,\l) \right|^2 
,\end{equation}
and agrees with the covariant calculation of $f_u(x)$ from its moments in Sec.~\ref{sec:DDcalc}. Similarly, the Dirac and Pauli
form factors can be expressed in terms of the effective wave function since the light-cone contour integration 
only encompasses the diquark pole. The expression are
\begin{eqnarray}
F_1(t) &=& \sum_{s_i} \int d  \mathbf{k}^\perp  \, dx \, \psi_{\text{LC}}^{*\text{eff}}(x, \mathbf{k}^\perp + (1  - x ) \mathbf{\D}^\perp; s_i, \l)
\psi_{\text{LC}}^{\text{eff}}(x, \mathbf{k}^\perp; s_i, \l), 
\\
- \frac{\l \D_\l} {2 M} F_2(t) &=& \sum_{s_i }\int d  \mathbf{k}^\perp  \, dx \, 
\psi_{\text{LC}}^{*\text{eff}}(x, \mathbf{k}^\perp + (1  - x ) \mathbf{\D}^\perp; s_i, -\l)
\psi_{\text{LC}}^{\text{eff}}(x, \mathbf{k}^\perp; s_i, \l) 
.\end{eqnarray}
Also for this reason, we can express the GPDs as simple convolutions in the diagonal overlap region $x>\x$. 
Defining $x_1 = \frac{x + \x}{1 + \x}$ and
$x_2 = \frac{ x - \x}{ 1 - \x}$, we have
\begin{multline}
\theta( x - \x) \mathcal{M}^{\l,\l}_q (x,\x,t)  = \sum_{s_i} \int d  \mathbf{k}^\perp  \, 
\psi_{\text{LC}}^{*\text{eff}} \left(x_2, \mathbf{k}^\perp + (1  - x_2 ) \frac{\mathbf{\D}^\perp}{2}; s_i, \l \right)
\\
\times
\psi_{\text{LC}}^{\text{eff}} \left(x_1, \mathbf{k}^\perp - (1  - x_1 ) \frac{\mathbf{\D}^\perp}{2}; s_i, \l \right) 
,\end{multline}
with a very similar expression holding for the spin-flip amplitude. 
In the above form the positivity bound is manifest.

In the axial-vector diquark channel we have the orthogonal amplitude
\begin{eqnarray} 
\psi (x,\mathbf{k}^\perp_{\text{rel}}; s_i,\l) 
&=& 
\frac{1}{(2 P^+)^2} \int \frac{d k^-}{2 \pi}
\frac{\ol u_{s_1} (P-k,m)}{\sqrt{1-x}} 
\Big( \gamma^+ i  \gamma_\mu  C \Big) 
\ol u_{s_2}(P-k,m)^{\text{T}} \notag \\ 
&& \times \frac{\ol u_{s_3} (k,m) }{\sqrt{x}} \gamma^+ \Psi^\mu(k,P) u_\l(P,M) 
.\label{eqn:lcwfn2} \end{eqnarray}
We again have omitted the color and isospin parts of the 
wave function which are: $\propto \varepsilon_{c_1,c_2,c_3} [ (\delta_{1,u} \delta_{2,d} + 
\delta_{1,d} \delta_{2,u} ) \delta_{3,u} - 2 \delta_{1,u} \delta_{2,u} \delta_{3,d}]$.
In this channel, the quark-diquark Bethe-Salpeter vector wave function is
\begin{equation}
\Psi^\mu(k,P) = \frac{i}{\rlap \slash k  - m + i \varepsilon} \Big ( - i g^{(a)} \gamma_\nu \gamma_5  \Big)  D^{\mu \nu}(P - k) 
,\end{equation} 
and we have chosen a single Dirac component $\Gamma^{(a)}_\nu = - i g^{(a)} \gamma_\nu \gamma_5$ 
and the coupling constant $g^{(a)} = g^{(s)}$. 
One can carry out the projection in Eq.~\eqref{eqn:lcwfn2} to find
analogous formulas for the axial diquark contribution to 
the quark distributions, form factors and GPDs for $x > \x$. 
The positivity bounds are again satisfied.

\section{Derivation of the double distributions} \label{sec:DDDD}

In this Appendix, we detail the calculation of the DDs in the scalar diquark channel and comment on the calculation in the 
axial-vector channel. The crucial steps in the derivation hinge upon reducing the numerator by factors present in the denominator
or by use of the simple identity:
\begin{equation} \label{eqn:notwogammas}
\gamma^{\{\mu_i} \gamma^{\mu_j\}} = 0
.\end{equation}

In the scalar diquark channel, let us take Eq.~\eqref{eqn:impulse} as our starting point. 
Denote the propagators simply by $\mathfrak{A} = (k - \ol P + \D /2)^2 - m_D^2 + i \varepsilon$, 
$\mathfrak{B} = (k + \D)^2 - m^2 + i \varepsilon$, and $\mathfrak{C} = k^2 - m^2 + i \varepsilon$.
The DDs in the scalar channel can be deduced without reducing factors in the numerator. We merely
introduce two Feynman parameters $\{x,y\}$ to cast the denominator specifically in the form
$[ x \mathfrak{A} + y \mathfrak{B} + (1-x-y) \mathfrak{C}]^{3}$. One then translates $k^\mu$ to
render the integral hyperspherically symmetric via the definition $k^\mu = l^\mu + \b \ol P {}^\mu - (\a + 1) \D^\mu / 2$.
Here $\b = x$ and $\a  = x + 2 y - 1$. Carrying out this procedure on Eq.~\eqref{eqn:impulse} produces
\begin{equation} \label{eqn:impulse2}
\Lambda_{ab} \int_0^1 d\b \int_{-1 + \b}^{1 - \b} d\a  \int d^4 l \,  
\, \ol u_{\l'}(\Pp)
\frac{N^{\{ \mu} (l + \b \ol P - \a \D /2)^{\mu_1} \cdots (l + \b \ol P - \a \D /2)^{\mu_n \} }   }
{[l^2 - D^{ab}(\b,\a;t)^{-1}]^{3}}
u_\l(P)
,\end{equation}   
where the numerator Dirac structure is given by
\begin{eqnarray} \label{eqn:firstN}
N^\mu &=& ( \lslash  + \b \Pslash - (\a - 1) \Dslash  / 2 ) \gamma^{\mu} ( \lslash  + \b  \Pslash  - (\a +1) \Dslash  / 2 )
\notag \\
&& + m^2 \gamma^\mu + i m \sigma^{\mu \nu} \D_\nu + 2 m ( l + \b \ol P - \a \D/ 2)^\mu 
.\end{eqnarray}

Using Eq.~\eqref{eqn:notwogammas}, as well as the identities in Appendix \ref{sec:gamma}, we can cast Eq.~\eqref{eqn:impulse2} in 
the form
\begin{multline} \label{eqn:impulse3}
\Lambda_{ab} \int_0^1 d\b \int_{-1 + \b}^{1 - \b} d\a  \int d^4 l  \;
[l^2 - D^{ab}(\b,\a;t)^{-1}]^{-3} \;
\ol u_{\l'}(\Pp)
\mathcal{N}^{\{\mu}  u_\l(P)
\\ \times
\sum_{k=0}^{n} \frac{n!}{ k! (n-k)!}     \b^{n-k}\a^k  
\ol P {}^{\mu_1} \cdots \ol P {}^{\mu_{n-k}} 
\left( - \frac{\D}{2}\right)^{\mu_{n-k+1}} \cdots \left( - \frac{\D}{2}\right)^{\mu_{n}\}}
,\end{multline}
where
\begin{eqnarray}
 \mathcal{N}^\mu &=& \left\{ (m + \b M)^2 - \frac{l^2}{2} + \frac{t}{4} [ (1-\b)^2 - \a^2]  \right\} \gamma^\mu  
\notag \\
&& + i \sigma^{\mu \nu} \D_\nu ( 1 - \b) ( m + \b M) - \a ( m + \b M) \D^\mu 
.\end{eqnarray}   
The $l$ integration is then standard and the DDs can be read off simply by using Eqs.~\eqref{eqn:moments} -- \eqref{eqn:generated}
and we arrive at Eq.~\eqref{eqn:Fs}.

Calculation in the axial-vector diquark channel is similar, however, the numerator is more complicated. In the same units as Eq.~\eqref{eqn:impulse}, 
we have the contribution from Fig.~\ref{ftwist} for the axial diquark
\begin{equation} \label{eqn:aximpulse}
\frac{\Lambda_{ab}}{9} Z(\Lambda) \int  d^4 k \, D^{\a \b}(k - P ) 
\frac{\ol u_{\l'} (\Pp) 
\gamma_5 \gamma_\a (\rlap\slash k + \Dslash  + m) \Gamma^{\mu \mu_1 \ldots \mu_n} (\rlap\slash k + m ) \gamma_\b \gamma_5  u_\l(P)}
{[k^2 - a^2 + i \varepsilon] [(k+\D)^2 - b^2 + i \varepsilon] }
,\end{equation}
where the axial-vector propagator, $D^{\a \b}(k - P)$, is given in Eq.~\eqref{eqn:vector}. The terms which result
from the $g^{\a \b}$ structure in the vector propagator can be dealt with straightforwardly after evaluating the contracted gamma
matrices. The integrals encountered are then similar to those in the scalar channel.

The terms in the numerator which arise from the second 
Lorentz structure of the vector propagator are more subtle. The Dirac structure of these terms is
\begin{equation} \label{eqn:secondN}
- \frac{1}{m_D^2}(\rlap \slash k  -  \Pslashe  ) 
[ - 2 m (k + \D/2)^\mu + m^2 \gamma^\mu - i m \sigma^{\mu \nu} \D_\nu + (\rlap \slash k +  \Dslash  ) \gamma^\mu  \rlap \slash k   ]
(\rlap \slash k  -   \Pslashe )
.\end{equation}
One must keep in mind that the index $\mu$ is entangled in the symmetric traceless combination
$\Gamma^{\mu \mu_1 \ldots \mu_n}$ given in Eq.~\eqref{eqn:GAMMA}. 
The first term in the square brackets of Eq.~\eqref{eqn:secondN} can be rewritten in the form
\begin{equation}
2 m (k + \D/2)^\mu (k  -  P )^2 = 2 m (k + \D/2)^\mu [\mathfrak{A} + m_D^2 ]   
,\end{equation}
where $\mathfrak{A}$ is the energy denominator of the spectator diquark and when canceled gives rise to a contribution $\propto \delta(\b)$ 
in $G(\b,\a;t)$. The remaining term $\propto m_D^2$ is simple to evaluate. The second term in Eq.~\eqref{eqn:secondN} 
is similarly easy to evaluate.

The remaining two terms' contributions to DDs are more involved. One must first reduce the quadruple and quintuple 
products of gamma matrices in these terms. After this procedure, one is left with terms not yet encountered above. 
After suitable algebra, these can be cast in the form\footnote{%
The case with an overall prefactor of $\D \cdot l$ proceeds as follows:
\begin{equation} \label{eqn:probterm2}
\D \cdot l \, ( l + \b \ol P - \a \D/ 2)^{\{ \mu} \cdots ( l + \b \ol P - \a \D/ 2)^{ \mu_{n-1}\}}
,\end{equation}
when contracted with lightlike vectors turns into
\begin{equation}
\frac{l^2}{4} \, n \, \D \cdot z \; ( \b \ol P \cdot z - \a \D \cdot z / 2)^{n-1} 
= 
- \frac{l^2}{2} \frac{\partial}{\partial \a} \sum_{k=0}^n \frac{n!}{k! (n-k)!}
(\b \ol P \cdot z)^{n-k} \left( -\a \D \cdot z/2 \right)^k 
.\end{equation}
In this form, we integrate by parts to read off contributions to the DDs, including the surface terms.}
\begin{equation} \label{eqn:probterm}
\ol P \cdot l \, ( l + \b \ol P - \a \D/ 2)^{\{ \mu} \cdots ( l + \b \ol P - \a \D/ 2)^{ \mu_{n-1}\}}
,\end{equation}
where we have shown only the relevant part of the numerator. 
Appealing to hyperspherical symmetry and contracting $n$ lightlike vectors, $z^{\mu_i}$, with the above expression, we have
\begin{equation}
\frac{l^2}{4} \, n \, \ol P \cdot z \; ( \b \ol P \cdot z - \a \D \cdot z / 2)^{n-1} = \frac{l^2}{4} \frac{\partial}{\partial \b} \sum_{k=0}^n \frac{n!}{k! (n-k)!}
(\b \ol P \cdot z)^{n-k} \left( -\a \D \cdot z/2 \right)^k 
.\end{equation}
Let us denote the result of the $l$-integration as $f(\b,\a;t)$.   
We have thus cast terms with the numerator Eq.~\eqref{eqn:probterm} in the form
\begin{multline}
\int_0^1 d\b \int_{-1 + \b}^{1 - \b} d\a \, f(\b,\a;t) \\
\times \frac{\partial}{\partial \b} \sum_{k=0}^n \frac{n!}{k! (n-k)!} \b^{n-k}\a^k 
\ol P {}^{\{\mu} \cdots \ol P {}^{\mu_{n-k-1}} 
\left( - \frac{\D}{2}\right)^{\mu_{n-k}} \cdots \left( - \frac{\D}{2}\right)^{\mu_{n-1}\}}
,\end{multline}
from which we change the order of integration and then integrate by parts to read off contributions to the DDs: 
$- \partial f(\b,\a;t) / \partial \b$, $-\delta(\b) f(\b,\a;t)$
and $\delta(\b - 1 + |\a|) f(\b,\a;t)$. The latter two are surface terms. In particular, 
the term proportional to $\delta(\b - 1 + |\a|)$ 
contributes to the DDs at their boundary of support and need not vanish \cite{Tiburzi:2004qr}.

Adding these contributions to those in Eq.~\eqref{eqn:Fa}, we arrive at the full expressions for the DDs in the 
axial diquark channel
{\tiny
\begin{eqnarray}
\begin{pmatrix}
F^{(a)}_q \\
K^{(a)}_q \\
G^{(a)}_q 
\end{pmatrix}
&=& \notag
\frac{2 N}{9} Z(\Lambda) (2 \delta_{q,d} + \delta_{q,u}) 
\left\{ 
\Lambda_{ab} D^{ab}
\begin{pmatrix}
m^2 + 5 \b m M + \frac{1}{2} \b [m_D^2 + M^2 (5 \b - 3) ] + \frac{t}{8} ( 5 \a^2 - \b^2 - 4 \b - 3)
- \frac{1}{4} [ (\b + \a) a^2 + (\b - \a) b^2]
\\
M \b  [ 3 m - 2 (1 -  \b) M]
\\
- 2 M  \a  [ 3 m + ( 1- 2 \b) M]
\end{pmatrix}
\right.
\\ \notag
&& + 
\Lambda_{ab} 
\log D^{ab}
\left[
1 + \frac{1}{2} 
[
\b \delta(|\a| + \b -1) 
+ 
(1-\b) \delta(\a + \b - 1)
+ 
(1-\b) \delta(\a - \b + 1) 
]
\right]
\begin{pmatrix}
1 \\
0 \\ 
0
\end{pmatrix} 
\\ \notag
&& + 
\frac{\L_{ab}}{2 m_D^2}
\left[
D^{ab}
\begin{pmatrix}
- m^2 ( M^2 (1 - \b)^2 + \frac{t}{4}[ (1 - \b)^2 - \a^2]) - \b t ( M^2 (1 - \b)^2 + \frac{t}{8} [ \a^2 - 2 (1 - \b)^2 ]) + M^2 (1 - \b) ( 2 M^2 \b - 2 m^2 + t)\\
2 m^2 M^2 (1 - \b )( 2 - \b)  + mM ( 2 M^2 ( 1 - \b)^2 + \a (a^2 - b^2 ) + [ (1 - \b)^2 - 3 \a^2] t / 2) + \b M^2 ( \frac{5}{2} \a^2 + 3 \b ( 2 - \b) - 2)\\
- 2 (1 - \b) M [ m (b^2 - a^2)  + \a (2 M^3 - 2 m^2 M + m t)] 
\end{pmatrix}
\right.
\\ \notag
&& \phantom{sp}
+
D^{ab}
\begin{pmatrix}
\frac{1}{4}[ a^2 + b^2 - 2 m_D^2 + 2 M^2 (1 - 2 \b) - (1 - \b) t ] ( - 2 m^2 + 2 \b M^2 + t [ 1 - 2 \b (1-\b)]) \\
2 \b M^2 ( b^2 ( \a - \b + 1 ) - a^2 ( \a + \b - 1) - (1-\b) [ 3 ( 2 \b - 1 ) m^2 + 2 m_D^2]) - M t\\
0
\end{pmatrix}
\\ \notag
&& \phantom{sp} + 
D^{ab}
[- m M t ( 1 - \b)^2 - \frac{\a}{2} ( M^2 + \b t) ( b^2 - a^2 + \a t)]
\begin{pmatrix} 
1 \\
0 \\
0
\end{pmatrix}
+
\log D^{ab} 
\begin{pmatrix}
2 M^2 + t (3 \b - 1) \\
2 M m - 4 M^2 (3 \b - 1)\\
0
\end{pmatrix}
\\ \notag
&& \phantom{sp} + 
\log D^{ab}  [\delta(\a + \b -1) - \delta(\a - \b + 1)]
\left.
\begin{pmatrix}
- \a ( M^2 + \b t)\\
2 \a M ( 2 \b M - m )\\
 - 4 m M ( 1 - \b )
\end{pmatrix}
+ \log D^{ab} [\delta(|\a| + \b - 1) - \delta(\b)]
\begin{pmatrix}
m^2 -\frac{t}{2} - \b M^2 + \b (1 - \b) t\\
- 4 M^2 \b (1- \b)\\
0
\end{pmatrix}
\right]
\\ \notag
&&  +
\frac{\L^2}{4 m_D^2}
\left[
\begin{pmatrix}
\left( \frac{m_D^2}{2} + \frac{\L^2}{4}+ (1-\b) \frac{t}{4} +  ( 2 \b - \frac{3}{2}) M^2 \right)(\L_a D^{a \L} + \L_b D^{\L b})
- \frac{3}{4} (\L_a D^{a \L} a^2  + \L_b D^{\L b} b^2)
+ \frac{\a t}{2}  (\L_a D^{a \L} - \L_b D^{\L b})  \\
(1-\b) M^2 (\L_a D^{a \L} + \L_b D^{\L b})\\
4 ( 1 - \b) M^2 (\L_a D^{a \L} - \L_b D^{\L b})
\end{pmatrix}
\right.
\\ \notag
&& \phantom{sp}
+
\frac{\L^2}{\L^2 - m^2} 
\left(
[
\delta(\a + \b - 1) - \delta(\a - \b + 1)
]
\log \frac{D^{m \L} }{D^{\L m}}
\left.
+ \frac{1}{2} 
[
\delta(|\a| + \b - 1) - \delta(\b)
]
\log \frac{D^{m \L} D^{\L m}}{(D^{\L \L})^2}
\right)
\begin{pmatrix}
1 \\
0 \\
0
\end{pmatrix} 
\right]
\\
&&+ 
\frac{1}{2 m_D^2}
\left[
\delta (\b) \Lambda_{ab} \log D^{ab}
\begin{pmatrix}
(1 + \a^2) \frac{t}{4} - 1 / D^{ab} \\
0 \\
4 \a M (M - m)
\end{pmatrix}
+
\left.
\frac{\delta(\b)}{2} \a (\L^2 - m^2) \log\frac{D^{\L m}}{D^{m \L}}
\begin{pmatrix}
1 \\
0 \\
0
\end{pmatrix} 
\right] 
\right\}
.\end{eqnarray}}
These cumbersome expressions belie the simplicity of calculating model DDs for the proton. 
Moreover the regularization scheme employed is likely inconsistent between the $g^{\mu \nu}$ and $k^\mu k^\nu$
terms. A particular signal of trouble is that the $d$-quark distribution calculated from the above 
expression becomes negative for small $x$. For these reasons, we have not presented the full results for 
the axial channel in the main text nor used them in our phenomenological analysis.
It is interesting to note that the largest contribution from the $k^\mu k^\nu$ piece is at small-$x$ and 
thus observables like the magnetic moment and charge radius are not terribly sensitive to the diagonal 
approximation.

\bibliography{lc.bib}

\begin{thebibliography}{53}
\expandafter\ifx\csname natexlab\endcsname\relax\def\natexlab#1{#1}\fi
\expandafter\ifx\csname bibnamefont\endcsname\relax
  \def\bibnamefont#1{#1}\fi
\expandafter\ifx\csname bibfnamefont\endcsname\relax
  \def\bibfnamefont#1{#1}\fi
\expandafter\ifx\csname citenamefont\endcsname\relax
  \def\citenamefont#1{#1}\fi
\expandafter\ifx\csname url\endcsname\relax
  \def\url#1{\texttt{#1}}\fi
\expandafter\ifx\csname urlprefix\endcsname\relax\def\urlprefix{URL }\fi
\providecommand{\bibinfo}[2]{#2}
\providecommand{\eprint}[2][]{\url{#2}}

\bibitem[{\citenamefont{M\"uller et~al.}(1994)\citenamefont{Muller, Robaschik,
  Geyer, Dittes, and Ho\v{r}ej\v{s}i}}]{Muller:1994fv}
\bibinfo{author}{\bibfnamefont{D.}~\bibnamefont{M\"uller}},
  \bibinfo{author}{\bibfnamefont{D.}~\bibnamefont{Robaschik}},
  \bibinfo{author}{\bibfnamefont{B.}~\bibnamefont{Geyer}},
  \bibinfo{author}{\bibfnamefont{F.~M.} \bibnamefont{Dittes}},
  \bibnamefont{and} \bibinfo{author}{\bibfnamefont{J.}~\bibnamefont{Ho\v{r}ej\v{s}i}},
  \bibinfo{journal}{Fortschr. Phys.} \textbf{\bibinfo{volume}{42}},
  \bibinfo{pages}{101} (\bibinfo{year}{1994}), \eprint{hep-ph/9812448}.

\bibitem[{\citenamefont{Radyushkin}(1996{\natexlab{a}})}]{Radyushkin:1996ru}
\bibinfo{author}{\bibfnamefont{A.~V.} \bibnamefont{Radyushkin}},
  \bibinfo{journal}{Phys. Lett.} \textbf{\bibinfo{volume}{B385}},
  \bibinfo{pages}{333} (\bibinfo{year}{1996}{\natexlab{a}}),
  \eprint{hep-ph/9605431}.

\bibitem[{\citenamefont{Radyushkin}(1996{\natexlab{b}})}]{Radyushkin:1996nd}
\bibinfo{author}{\bibfnamefont{A.~V.} \bibnamefont{Radyushkin}},
  \bibinfo{journal}{Phys. Lett.} \textbf{\bibinfo{volume}{B380}},
  \bibinfo{pages}{417} (\bibinfo{year}{1996}{\natexlab{b}}),
  \eprint{hep-ph/9604317}.

\bibitem[{\citenamefont{Ji}(1997{\natexlab{a}})}]{Ji:1997ek}
\bibinfo{author}{\bibfnamefont{X.-D.} \bibnamefont{Ji}},
  \bibinfo{journal}{Phys. Rev. Lett.} \textbf{\bibinfo{volume}{78}},
  \bibinfo{pages}{610} (\bibinfo{year}{1997}{\natexlab{a}}),
  \eprint{hep-ph/9603249}.

\bibitem[{\citenamefont{Ji}(1997{\natexlab{b}})}]{Ji:1997nm}
\bibinfo{author}{\bibfnamefont{X.-D.} \bibnamefont{Ji}},
  \bibinfo{journal}{Phys. Rev.} \textbf{\bibinfo{volume}{D55}},
  \bibinfo{pages}{7114} (\bibinfo{year}{1997}{\natexlab{b}}),
  \eprint{hep-ph/9609381}.

\bibitem[{\citenamefont{Ji}(1998)}]{Ji:1998pc}
\bibinfo{author}{\bibfnamefont{X.-D.} \bibnamefont{Ji}}, \bibinfo{journal}{J.
  Phys.} \textbf{\bibinfo{volume}{G24}}, \bibinfo{pages}{1181}
  (\bibinfo{year}{1998}), \eprint{hep-ph/9807358}.

\bibitem[{\citenamefont{Radyushkin}(2000)}]{Radyushkin:2000uy}
\bibinfo{author}{\bibfnamefont{A.~V.} \bibnamefont{Radyushkin}}
  (\bibinfo{year}{2000}), \eprint{hep-ph/0101225}.

\bibitem[{\citenamefont{Goeke et~al.}(2001)\citenamefont{Goeke, Polyakov, and
  Vanderhaeghen}}]{Goeke:2001tz}
\bibinfo{author}{\bibfnamefont{K.}~\bibnamefont{Goeke}},
  \bibinfo{author}{\bibfnamefont{M.~V.} \bibnamefont{Polyakov}},
  \bibnamefont{and}
  \bibinfo{author}{\bibfnamefont{M.}~\bibnamefont{Vanderhaeghen}},
  \bibinfo{journal}{Prog. Part. Nucl. Phys.} \textbf{\bibinfo{volume}{47}},
  \bibinfo{pages}{401} (\bibinfo{year}{2001}), \eprint{hep-ph/0106012}.

\bibitem[{\citenamefont{Belitsky et~al.}(2002)\citenamefont{Belitsky, M\"uller,
  and Kirchner}}]{Belitsky:2001ns}
\bibinfo{author}{\bibfnamefont{A.~V.} \bibnamefont{Belitsky}},
  \bibinfo{author}{\bibfnamefont{D.}~\bibnamefont{M\"uller}}, \bibnamefont{and}
  \bibinfo{author}{\bibfnamefont{A.}~\bibnamefont{Kirchner}},
  \bibinfo{journal}{Nucl. Phys.} \textbf{\bibinfo{volume}{B629}},
  \bibinfo{pages}{323} (\bibinfo{year}{2002}), \eprint{hep-ph/0112108}.

\bibitem[{\citenamefont{Diehl}(2003)}]{Diehl:2003ny}
\bibinfo{author}{\bibfnamefont{M.}~\bibnamefont{Diehl}},
  \bibinfo{journal}{Phys. Rept.} \textbf{\bibinfo{volume}{388}},
  \bibinfo{pages}{41} (\bibinfo{year}{2003}), \eprint{hep-ph/0307382}.

\bibitem[{\citenamefont{Diehl et~al.}(2001)\citenamefont{Diehl, Feldmann,
  Jakob, and Kroll}}]{Diehl:2000xz}
\bibinfo{author}{\bibfnamefont{M.}~\bibnamefont{Diehl}},
  \bibinfo{author}{\bibfnamefont{T.}~\bibnamefont{Feldmann}},
  \bibinfo{author}{\bibfnamefont{R.}~\bibnamefont{Jakob}}, \bibnamefont{and}
  \bibinfo{author}{\bibfnamefont{P.}~\bibnamefont{Kroll}},
  \bibinfo{journal}{Nucl. Phys.} \textbf{\bibinfo{volume}{B596}},
  \bibinfo{pages}{33} (\bibinfo{year}{2001}), \eprint{hep-ph/0009255}.

\bibitem[{\citenamefont{Brodsky et~al.}(2001)\citenamefont{Brodsky, Diehl, and
  Hwang}}]{Brodsky:2000xy}
\bibinfo{author}{\bibfnamefont{S.~J.} \bibnamefont{Brodsky}},
  \bibinfo{author}{\bibfnamefont{M.}~\bibnamefont{Diehl}}, \bibnamefont{and}
  \bibinfo{author}{\bibfnamefont{D.~S.} \bibnamefont{Hwang}},
  \bibinfo{journal}{Nucl. Phys.} \textbf{\bibinfo{volume}{B596}},
  \bibinfo{pages}{99} (\bibinfo{year}{2001}), \eprint{hep-ph/0009254}.

\bibitem[{\citenamefont{Burkardt}(2000)}]{Burkardt:2000za}
\bibinfo{author}{\bibfnamefont{M.}~\bibnamefont{Burkardt}},
  \bibinfo{journal}{Phys. Rev.} \textbf{\bibinfo{volume}{D62}},
  \bibinfo{pages}{071503} (\bibinfo{year}{2000}), \eprint{hep-ph/0005108}.

\bibitem[{\citenamefont{Diehl}(2002)}]{Diehl:2002he}
\bibinfo{author}{\bibfnamefont{M.}~\bibnamefont{Diehl}}, \bibinfo{journal}{Eur.
  Phys. J.} \textbf{\bibinfo{volume}{C25}}, \bibinfo{pages}{223}
  (\bibinfo{year}{2002}), \eprint{hep-ph/0205208}.

\bibitem[{\citenamefont{Burkardt}(2003)}]{Burkardt:2002hr}
\bibinfo{author}{\bibfnamefont{M.}~\bibnamefont{Burkardt}},
  \bibinfo{journal}{Int. J. Mod. Phys.} \textbf{\bibinfo{volume}{A18}},
  \bibinfo{pages}{173} (\bibinfo{year}{2003}), \eprint{hep-ph/0207047}.

\bibitem[{\citenamefont{Radyushkin}(1999)}]{Radyushkin:1998es}
\bibinfo{author}{\bibfnamefont{A.~V.} \bibnamefont{Radyushkin}},
  \bibinfo{journal}{Phys. Rev.} \textbf{\bibinfo{volume}{D59}},
  \bibinfo{pages}{014030} (\bibinfo{year}{1999}), \eprint{hep-ph/9805342}.

\bibitem[{\citenamefont{Pire et~al.}(1999)\citenamefont{Pire, Soffer, and
  Teryaev}}]{Pire:1998nw}
\bibinfo{author}{\bibfnamefont{B.}~\bibnamefont{Pire}},
  \bibinfo{author}{\bibfnamefont{J.}~\bibnamefont{Soffer}}, \bibnamefont{and}
  \bibinfo{author}{\bibfnamefont{O.}~\bibnamefont{Teryaev}},
  \bibinfo{journal}{Eur. Phys. J.} \textbf{\bibinfo{volume}{C8}},
  \bibinfo{pages}{103} (\bibinfo{year}{1999}), \eprint{hep-ph/9804284}.

\bibitem[{\citenamefont{Pobylitsa}(2002{\natexlab{a}})}]{Pobylitsa:2001nt}
\bibinfo{author}{\bibfnamefont{P.~V.} \bibnamefont{Pobylitsa}},
  \bibinfo{journal}{Phys. Rev.} \textbf{\bibinfo{volume}{D65}},
  \bibinfo{pages}{077504} (\bibinfo{year}{2002}{\natexlab{a}}),
  \eprint{hep-ph/0112322}.

\bibitem[{\citenamefont{Pobylitsa}(2002{\natexlab{b}})}]{Pobylitsa:2002gw}
\bibinfo{author}{\bibfnamefont{P.~V.} \bibnamefont{Pobylitsa}},
  \bibinfo{journal}{Phys. Rev.} \textbf{\bibinfo{volume}{D65}},
  \bibinfo{pages}{114015} (\bibinfo{year}{2002}{\natexlab{b}}),
  \eprint{hep-ph/0201030}.

\bibitem[{\citenamefont{Pobylitsa}(2002{\natexlab{c}})}]{Pobylitsa:2002iu}
\bibinfo{author}{\bibfnamefont{P.~V.} \bibnamefont{Pobylitsa}},
  \bibinfo{journal}{Phys. Rev.} \textbf{\bibinfo{volume}{D66}},
  \bibinfo{pages}{094002} (\bibinfo{year}{2002}{\natexlab{c}}),
  \eprint{hep-ph/0204337}.

\bibitem[{\citenamefont{Pobylitsa}(2004)}]{Pobylitsa:2002ru}
\bibinfo{author}{\bibfnamefont{P.~V.} \bibnamefont{Pobylitsa}},
  \bibinfo{journal}{Phys. Rev.} \textbf{\bibinfo{volume}{D70}},
  \bibinfo{pages}{034004} (\bibinfo{year}{2004}), \eprint{hep-ph/0211160}.

\bibitem[{\citenamefont{Tiburzi and
  Miller}(2003{\natexlab{a}})}]{Tiburzi:2002sx}
\bibinfo{author}{\bibfnamefont{B.~C.} \bibnamefont{Tiburzi}} \bibnamefont{and}
  \bibinfo{author}{\bibfnamefont{G.~A.} \bibnamefont{Miller}},
  \bibinfo{journal}{Phys. Rev.} \textbf{\bibinfo{volume}{D67}},
  \bibinfo{pages}{054015} (\bibinfo{year}{2003}{\natexlab{a}}),
  \eprint{hep-ph/0210305}.

\bibitem[{\citenamefont{Radyushkin}(1997)}]{Radyushkin:1997ki}
\bibinfo{author}{\bibfnamefont{A.~V.} \bibnamefont{Radyushkin}},
  \bibinfo{journal}{Phys. Rev.} \textbf{\bibinfo{volume}{D56}},
  \bibinfo{pages}{5524} (\bibinfo{year}{1997}), \eprint{hep-ph/9704207}.

\bibitem[{\citenamefont{Mukherjee et~al.}(2003)\citenamefont{Mukherjee,
  Musatov, Pauli, and Radyushkin}}]{Mukherjee:2002gb}
\bibinfo{author}{\bibfnamefont{A.}~\bibnamefont{Mukherjee}},
  \bibinfo{author}{\bibfnamefont{I.~V.} \bibnamefont{Musatov}},
  \bibinfo{author}{\bibfnamefont{H.~C.} \bibnamefont{Pauli}}, \bibnamefont{and}
  \bibinfo{author}{\bibfnamefont{A.~V.} \bibnamefont{Radyushkin}},
  \bibinfo{journal}{Phys. Rev.} \textbf{\bibinfo{volume}{D67}},
  \bibinfo{pages}{073014} (\bibinfo{year}{2003}), \eprint{hep-ph/0205315}.

\bibitem[{\citenamefont{Pobylitsa}(2003{\natexlab{a}})}]{Pobylitsa:2002vw}
\bibinfo{author}{\bibfnamefont{P.~V.} \bibnamefont{Pobylitsa}},
  \bibinfo{journal}{Phys. Rev.} \textbf{\bibinfo{volume}{D67}},
  \bibinfo{pages}{094012} (\bibinfo{year}{2003}{\natexlab{a}}),
  \eprint{hep-ph/0210238}.

\bibitem[{\citenamefont{Pobylitsa}(2003{\natexlab{b}})}]{Pobylitsa:2002vi}
\bibinfo{author}{\bibfnamefont{P.~V.} \bibnamefont{Pobylitsa}},
  \bibinfo{journal}{Phys. Rev.} \textbf{\bibinfo{volume}{D67}},
  \bibinfo{pages}{034009} (\bibinfo{year}{2003}{\natexlab{b}}),
  \eprint{hep-ph/0210150}.

\bibitem[{\citenamefont{Tiburzi and
  Miller}(2003{\natexlab{b}})}]{Tiburzi:2002kr}
\bibinfo{author}{\bibfnamefont{B.~C.} \bibnamefont{Tiburzi}} \bibnamefont{and}
  \bibinfo{author}{\bibfnamefont{G.~A.} \bibnamefont{Miller}},
  \bibinfo{journal}{Phys. Rev.} \textbf{\bibinfo{volume}{D67}},
  \bibinfo{pages}{013010} (\bibinfo{year}{2003}{\natexlab{b}}),
  \eprint{hep-ph/0209178}.

\bibitem[{\citenamefont{Tiburzi and
  Miller}(2003{\natexlab{c}})}]{Tiburzi:2002tq}
\bibinfo{author}{\bibfnamefont{B.~C.} \bibnamefont{Tiburzi}} \bibnamefont{and}
  \bibinfo{author}{\bibfnamefont{G.~A.} \bibnamefont{Miller}},
  \bibinfo{journal}{Phys. Rev.} \textbf{\bibinfo{volume}{D67}},
  \bibinfo{pages}{113004} (\bibinfo{year}{2003}{\natexlab{c}}),
  \eprint{hep-ph/0212238}.

\bibitem[{\citenamefont{Tiburzi et~al.}(2003)\citenamefont{Tiburzi, Detmold,
  and Miller}}]{Tiburzi:2003ja}
\bibinfo{author}{\bibfnamefont{B.~C.} \bibnamefont{Tiburzi}},
  \bibinfo{author}{\bibfnamefont{W.}~\bibnamefont{Detmold}}, \bibnamefont{and}
  \bibinfo{author}{\bibfnamefont{G.~A.} \bibnamefont{Miller}},
  \bibinfo{journal}{Phys. Rev.} \textbf{\bibinfo{volume}{D68}},
  \bibinfo{pages}{073002} (\bibinfo{year}{2003}), \eprint{hep-ph/0305190}.

\bibitem[{\citenamefont{Tiburzi}(2004{\natexlab{a}})}]{Tiburzi:2004ye}
\bibinfo{author}{\bibfnamefont{B.~C.} \bibnamefont{Tiburzi}},
\bibinfo{journal}{Ph.D.~thesis, University of Washington}
  (\bibinfo{year}{2004}{\natexlab{a}}), \eprint{nucl-th/0407005}.

\bibitem[{\citenamefont{Polyakov and Weiss}(1999)}]{Polyakov:1999gs}
\bibinfo{author}{\bibfnamefont{M.~V.} \bibnamefont{Polyakov}} \bibnamefont{and}
  \bibinfo{author}{\bibfnamefont{C.}~\bibnamefont{Weiss}},
  \bibinfo{journal}{Phys. Rev.} \textbf{\bibinfo{volume}{D60}},
  \bibinfo{pages}{114017} (\bibinfo{year}{1999}), \eprint{hep-ph/9902451}.

\bibitem[{\citenamefont{Belitsky et~al.}(2001)\citenamefont{Belitsky, M\"uller,
  Kirchner, and Sch\"afer}}]{Belitsky:2000vk}
\bibinfo{author}{\bibfnamefont{A.~V.} \bibnamefont{Belitsky}},
  \bibinfo{author}{\bibfnamefont{D.}~\bibnamefont{M\"uller}},
  \bibinfo{author}{\bibfnamefont{A.}~\bibnamefont{Kirchner}}, \bibnamefont{and}
  \bibinfo{author}{\bibfnamefont{A.}~\bibnamefont{Sch\"afer}},
  \bibinfo{journal}{Phys. Rev.} \textbf{\bibinfo{volume}{D64}},
  \bibinfo{pages}{116002} (\bibinfo{year}{2001}), \eprint{hep-ph/0011314}.

\bibitem[{\citenamefont{Teryaev}(2001)}]{Teryaev:2001qm}
\bibinfo{author}{\bibfnamefont{O.~V.} \bibnamefont{Teryaev}},
  \bibinfo{journal}{Phys. Lett.} \textbf{\bibinfo{volume}{B510}},
  \bibinfo{pages}{125} (\bibinfo{year}{2001}), \eprint{hep-ph/0102303}.

\bibitem[{\citenamefont{Tiburzi}(2004{\natexlab{b}})}]{Tiburzi:2004qr}
\bibinfo{author}{\bibfnamefont{B.~C.} \bibnamefont{Tiburzi}}
  (\bibinfo{year}{2004}{\natexlab{b}}), \eprint{hep-ph/0405211}.

\bibitem[{\citenamefont{Chung and Coester}(1991)}]{Chung:1991st}
\bibinfo{author}{\bibfnamefont{P.~L.} \bibnamefont{Chung}} \bibnamefont{and}
  \bibinfo{author}{\bibfnamefont{F.}~\bibnamefont{Coester}},
  \bibinfo{journal}{Phys. Rev.} \textbf{\bibinfo{volume}{D44}},
  \bibinfo{pages}{229} (\bibinfo{year}{1991}).

\bibitem[{\citenamefont{Buck et~al.}(1992)\citenamefont{Buck, Alkofer, and
  Reinhardt}}]{Buck:1992wz}
\bibinfo{author}{\bibfnamefont{A.}~\bibnamefont{Buck}},
  \bibinfo{author}{\bibfnamefont{R.}~\bibnamefont{Alkofer}}, \bibnamefont{and}
  \bibinfo{author}{\bibfnamefont{H.}~\bibnamefont{Reinhardt}},
  \bibinfo{journal}{Phys. Lett.} \textbf{\bibinfo{volume}{B286}},
  \bibinfo{pages}{29} (\bibinfo{year}{1992}).

\bibitem[{\citenamefont{Mineo et~al.}(1999)\citenamefont{Mineo, Bentz, and
  Yazaki}}]{Mineo:1999eq}
\bibinfo{author}{\bibfnamefont{H.}~\bibnamefont{Mineo}},
  \bibinfo{author}{\bibfnamefont{W.}~\bibnamefont{Bentz}}, \bibnamefont{and}
  \bibinfo{author}{\bibfnamefont{K.}~\bibnamefont{Yazaki}},
  \bibinfo{journal}{Phys. Rev.} \textbf{\bibinfo{volume}{C60}},
  \bibinfo{pages}{065201} (\bibinfo{year}{1999}), \eprint{nucl-th/9907043}.

\bibitem[{\citenamefont{Oettel et~al.}(1998)\citenamefont{Oettel, Hellstern,
  Alkofer, and Reinhardt}}]{Oettel:1998bk}
\bibinfo{author}{\bibfnamefont{M.}~\bibnamefont{Oettel}},
  \bibinfo{author}{\bibfnamefont{G.}~\bibnamefont{Hellstern}},
  \bibinfo{author}{\bibfnamefont{R.}~\bibnamefont{Alkofer}}, \bibnamefont{and}
  \bibinfo{author}{\bibfnamefont{H.}~\bibnamefont{Reinhardt}},
  \bibinfo{journal}{Phys. Rev.} \textbf{\bibinfo{volume}{C58}},
  \bibinfo{pages}{2459} (\bibinfo{year}{1998}), \eprint{nucl-th/9805054}.


\bibitem[{\citenamefont{Oettel et~al.}(2000)\citenamefont{Oettel,
  Alkofer, and von Smekal}}]{Oettel:2000jj}
\bibinfo{author}{\bibfnamefont{M.}~\bibnamefont{Oettel}},
  \bibinfo{author}{\bibfnamefont{R.}~\bibnamefont{Alkofer}}, \bibnamefont{and}
\bibinfo{author}{\bibfnamefont{L.}~\bibnamefont{von Smekal}},
 \bibinfo{journal}{Eur. Phys. J.} \textbf{\bibinfo{volume}{A8}},
  \bibinfo{pages}{553} (\bibinfo{year}{2000}), \eprint{nucl-th/0006082}.

\bibitem[{\citenamefont{Oettel}(2000)}]{Oettel:2000ig}
\bibinfo{author}{\bibfnamefont{M.}~\bibnamefont{Oettel}},
\bibinfo{journal}{Ph.D.~thesis, University of T\"ubingen}
  (\bibinfo{year}{2000}), \eprint{nucl-th/0012067}.

\bibitem[{\citenamefont{Theussl et~al.}(2004)\citenamefont{Theussl, Noguera,
  and Vento}}]{Theussl:2002xp}
\bibinfo{author}{\bibfnamefont{L.}~\bibnamefont{Theussl}},
  \bibinfo{author}{\bibfnamefont{S.}~\bibnamefont{Noguera}}, \bibnamefont{and}
  \bibinfo{author}{\bibfnamefont{V.}~\bibnamefont{Vento}},
  \bibinfo{journal}{Eur. Phys. J.} \textbf{\bibinfo{volume}{A20}},
  \bibinfo{pages}{483} (\bibinfo{year}{2004}), \eprint{nucl-th/0211036}.

\bibitem[{\citenamefont{Bakker et~al.}(2001)\citenamefont{Bakker, Choi, and
  Ji}}]{Bakker:2000pk}
\bibinfo{author}{\bibfnamefont{B.~L.~G.} \bibnamefont{Bakker}},
  \bibinfo{author}{\bibfnamefont{H.-M.} \bibnamefont{Choi}}, \bibnamefont{and}
  \bibinfo{author}{\bibfnamefont{C.-R.} \bibnamefont{Ji}},
  \bibinfo{journal}{Phys. Rev.} \textbf{\bibinfo{volume}{D63}},
  \bibinfo{pages}{074014} (\bibinfo{year}{2001}), \eprint{hep-ph/0008147}.

\bibitem[{\citenamefont{Jones et~al.}(2000)}]{Jones:1999rz}
\bibinfo{author}{\bibfnamefont{M.~K.} \bibnamefont{Jones}} \bibnamefont{et~al.}
  (\bibinfo{collaboration}{Jefferson Lab Hall A}), \bibinfo{journal}{Phys. Rev.
  Lett.} \textbf{\bibinfo{volume}{84}}, \bibinfo{pages}{1398}
  (\bibinfo{year}{2000}), \eprint{nucl-ex/9910005}.

\bibitem[{\citenamefont{Gayou et~al.}(2002)}]{Gayou:2001qd}
\bibinfo{author}{\bibfnamefont{O.}~\bibnamefont{Gayou}} \bibnamefont{et~al.}
  (\bibinfo{collaboration}{Jefferson Lab Hall A}), \bibinfo{journal}{Phys. Rev.
  Lett.} \textbf{\bibinfo{volume}{88}}, \bibinfo{pages}{092301}
  (\bibinfo{year}{2002}), \eprint{nucl-ex/0111010}.

\bibitem[{\citenamefont{Arrington}(2004)}]{Arrington:2003qk}
\bibinfo{author}{\bibfnamefont{J.}~\bibnamefont{Arrington}},
  \bibinfo{journal}{Phys. Rev.} \textbf{\bibinfo{volume}{C69}},
  \bibinfo{pages}{022201} (\bibinfo{year}{2004}), \eprint{nucl-ex/0309011}.

\bibitem[{\citenamefont{Belitsky and
  M\"uller}(1999{\natexlab{a}})}]{Belitsky:1998gc}
\bibinfo{author}{\bibfnamefont{A.~V.} \bibnamefont{Belitsky}} \bibnamefont{and}
  \bibinfo{author}{\bibfnamefont{D.}~\bibnamefont{M\"uller}},
  \bibinfo{journal}{Nucl. Phys.} \textbf{\bibinfo{volume}{B537}},
  \bibinfo{pages}{397} (\bibinfo{year}{1999}{\natexlab{a}}),
  \eprint{hep-ph/9804379}.

\bibitem[{\citenamefont{Belitsky
  et~al.}(1999{\natexlab{a}})\citenamefont{Belitsky, M\"uller, Niedermeier, and
  Sch\"afer}}]{Belitsky:1998uk}
\bibinfo{author}{\bibfnamefont{A.~V.} \bibnamefont{Belitsky}},
  \bibinfo{author}{\bibfnamefont{D.}~\bibnamefont{M\"uller}},
  \bibinfo{author}{\bibfnamefont{L.}~\bibnamefont{Niedermeier}},
  \bibnamefont{and} \bibinfo{author}{\bibfnamefont{A.}~\bibnamefont{Sch\"afer}},
  \bibinfo{journal}{Nucl. Phys.} \textbf{\bibinfo{volume}{B546}},
  \bibinfo{pages}{279} (\bibinfo{year}{1999}{\natexlab{a}}),
  \eprint{hep-ph/9810275}.

\bibitem[{\citenamefont{Belitsky
  et~al.}(1999{\natexlab{b}})\citenamefont{Belitsky, M\"uller, and
  Freund}}]{Belitsky:1999gu}
\bibinfo{author}{\bibfnamefont{A.~V.} \bibnamefont{Belitsky}},
  \bibinfo{author}{\bibfnamefont{D.}~\bibnamefont{M\"uller}}, \bibnamefont{and}
  \bibinfo{author}{\bibfnamefont{A.}~\bibnamefont{Freund}},
  \bibinfo{journal}{Phys. Lett.} \textbf{\bibinfo{volume}{B461}},
  \bibinfo{pages}{270} (\bibinfo{year}{1999}{\natexlab{b}}),
  \eprint{hep-ph/9904477}.

\bibitem[{\citenamefont{Belitsky and
  M\"uller}(1999{\natexlab{b}})}]{Belitsky:1999fu}
\bibinfo{author}{\bibfnamefont{A.~V.} \bibnamefont{Belitsky}} \bibnamefont{and}
  \bibinfo{author}{\bibfnamefont{D.}~\bibnamefont{M\"uller}},
  \bibinfo{journal}{Phys. Lett.} \textbf{\bibinfo{volume}{B464}},
  \bibinfo{pages}{249} (\bibinfo{year}{1999}{\natexlab{b}}),
  \eprint{hep-ph/9906409}.

\bibitem[{\citenamefont{Belitsky et~al.}(2000)\citenamefont{Belitsky, Freund,
  and M\"uller}}]{Belitsky:1999hf}
\bibinfo{author}{\bibfnamefont{A.~V.} \bibnamefont{Belitsky}},
  \bibinfo{author}{\bibfnamefont{A.}~\bibnamefont{Freund}}, \bibnamefont{and}
  \bibinfo{author}{\bibfnamefont{D.}~\bibnamefont{M\"uller}},
  \bibinfo{journal}{Nucl. Phys.} \textbf{\bibinfo{volume}{B574}},
  \bibinfo{pages}{347} (\bibinfo{year}{2000}), \eprint{hep-ph/9912379}.

\bibitem[{\citenamefont{Brown}(1992)}]{Brown:1992db}
\bibinfo{author}{\bibfnamefont{L.~S.} \bibnamefont{Brown}},
  \emph{\bibinfo{title}{Quantum field theory}} (\bibinfo{publisher}{Cambridge,
  UK: Univ. Pr.}, \bibinfo{year}{1992}).

\bibitem[{\citenamefont{Heinzl}(1998)}]{Heinzl:1998kz}
\bibinfo{author}{\bibfnamefont{T.}~\bibnamefont{Heinzl}}
  (\bibinfo{year}{1998}), \eprint{hep-th/9812190}.

\bibitem[{\citenamefont{Ji et~al.}(2003)\citenamefont{Ji, Ma, and
  Yuan}}]{Ji:2002xn}
\bibinfo{author}{\bibfnamefont{X.-D.} \bibnamefont{Ji}},
  \bibinfo{author}{\bibfnamefont{J.-P.} \bibnamefont{Ma}}, \bibnamefont{and}
  \bibinfo{author}{\bibfnamefont{F.}~\bibnamefont{Yuan}},
  \bibinfo{journal}{Nucl. Phys.} \textbf{\bibinfo{volume}{B652}},
  \bibinfo{pages}{383} (\bibinfo{year}{2003}), \eprint{hep-ph/0210430}.

\bibitem[{\citenamefont{Tiburzi and
  Miller}(2003{\natexlab{d}})}]{Tiburzi:2002sw}
\bibinfo{author}{\bibfnamefont{B.~C.} \bibnamefont{Tiburzi}} \bibnamefont{and}
  \bibinfo{author}{\bibfnamefont{G.~A.} \bibnamefont{Miller}},
  \bibinfo{journal}{Phys. Rev.} \textbf{\bibinfo{volume}{D67}},
  \bibinfo{pages}{054014} (\bibinfo{year}{2003}{\natexlab{d}}),
  \eprint{hep-ph/0210304}.

\end{thebibliography}

\end{document}